\documentclass[aps,prd,nofootinbib,showpacs,superscriptaddress,twocolumn,floatfix]{revtex4-1}

\usepackage{amsmath,amssymb}
\usepackage[utf8]{inputenc}
\usepackage{graphicx}
\usepackage{mathrsfs}
\usepackage{bm}
\usepackage{bbm}
\usepackage{indentfirst}
\usepackage{epstopdf}
\usepackage{color}
\usepackage{xcolor}
\usepackage{amssymb, amsmath,color, hyperref, graphicx}

\usepackage{braket}
\usepackage{placeins}
\usepackage{multirow}
\usepackage{slashed}
\usepackage{physics}

\newcommand{\bec}{\begin{center}}
\newcommand{\eec}{\end{center}}
\newcommand{\beq}{\begin{equation}}
\newcommand{\eeq}{\end{equation}}
\newcommand{\bea}{\begin{eqnarray}}
\newcommand{\eea}{\end{eqnarray}}
\newcommand{\nn}{\nonumber}

\newcommand{\tB}{{\rm B}}
\newcommand{\tQ}{{\rm Q}}
\newcommand{\tS}{{\rm S}}

\newcommand{\hmu}{{\hat{\mu}}}

\newcommand{\muB}{{{\mu}_{\rm B}}}

\begin{document}

\title{Isospin-Driven Splitting of Chemical Potentials in Isobar Collisions from Lattice QCD}

\author{Heng-Tong Ding} 
\author{Jin-Biao Gu} 
\author{Arpith Kumar} 
\author{Jia Ni} 

\affiliation{Key Laboratory of Quark and Lepton Physics (MOE) and Institute of
Particle Physics, Central China Normal University, Wuhan 430079, China}

\date{\today}
\begin{abstract}

Strong magnetic fields produced in relativistic heavy-ion collisions can modify fluctuations of conserved charges and, consequently, their associated chemical potentials. We present first-principles $(2+1)$-flavor lattice-QCD results for isospin-driven splittings of conserved-charge chemical potentials between the isobar systems $^{96}_{44}\mathrm{Ru}+^{96}_{44}\mathrm{Ru}$ and $^{96}_{40}\mathrm{Zr}+^{96}_{40}\mathrm{Zr}$ in the QCD crossover region, both at vanishing and nonzero magnetic fields along the pseudo-critical line $T_{pc}(eB)$. We outline a framework that, under strangeness neutrality and charge-to-baryon ratio $r\equiv n_\tQ/n_\tB$, maps the isospin difference between two nuclei, as encoded in $r_{\rm Zr}$ and $r_{\rm Ru}$, onto splitting ratios $\Delta\mu_\tQ/\Delta\mu_\tB$, $\Delta\mu_\tS/\Delta\mu_\tB$, and $\Delta\mu_\tS/\Delta\mu_\tQ$ as functions of $\mu_\tB(r_{\rm Ru})/\Delta\mu_\tB$. 
Using continuum-estimated lattice results for the leading-order coefficients $q_1\equiv(\mu_\tQ/\mu_\tB)_{\rm LO}$ and $s_1\equiv(\mu_\tS/\mu_\tB)_{\rm LO}$, we find that, at vanishing magnetic field, the splitting ratios are of similar magnitude to recent Bayesian extractions from STAR isobar data and yield $\Delta\mu_\tQ<0$ and $\Delta\mu_\tS>0$, with the electric-charge sector dominating. At nonzero magnetic fields, the splitting ratios show only moderate $eB$ dependence. We therefore further examine Ru--Zr differences in the normalized magnetic-field response of chemical-potential ratios, particularly those involving $\mu_\tQ/\mu_\tB$, which display a pronounced enhancement in lattice QCD. We also present hadron resonance gas (HRG) results and experimentally motivated proxy observables with kinematic cuts to facilitate contact with experiment.
  
\end{abstract}


\maketitle

\section{Introduction}
\label{sec:intro}

Strong magnetic fields of order $eB\sim\Lambda_{\rm QCD}^2$ are expected in off-central relativistic heavy-ion collisions~\cite{Kharzeev:2007jp,Skokov:2009qp,Deng:2012pc}. At early stages, model estimates suggest  $eB\sim5~M_\pi^2$ at RHIC and $eB\sim70~M_\pi^2$ at LHC for $^{197}_{79}$Au$\big/ ~^{208}_{82}$Pb nuclei collisions~\cite{Deng:2012pc,Skokov:2009qp}. Although transient,
a sufficiently sustained lifetime---supported by the medium's electrical conductivity and paramagnetic properties~\cite{Astrakhantsev:2019zkr,Bali:2020bcn,Ding:2016hua} governed by magnetohydrodynamics~\cite{Huang:2022qdn}---could allow these fields to induce notable non-perturbative effects on the produced QGP, such as the chiral magnetic effect~\cite{Kharzeev:2007jp,Fukushima:2008xe,Kharzeev:2012ph,Kharzeev:2020jxw}. This opens the possibility for their detection through final-state observables, motivating intensive theoretical and experimental investigations~\cite{Fukushima:2009ft,Fu:2013ica, Fukushima:2016vix,STAR:2021mii,Kharzeev:2022hqz,Ding:2023bft,Brandt:2024fpc,ALICE:2025mkk} (see~\cite{Endrodi:2024cqn,Adhikari:2024bfa,Ding:2026gao,Brandt:2026dcd} for recent reviews).

Over the past decade, lattice-QCD studies in magnetic backgrounds have revealed significant modifications in QCD thermodynamics, including isospin symmetry breaking, reduction of the transition temperature and inverse magnetic catalysis~\cite{Bali:2011qj,Bali:2012zg,DElia:2011koc,DElia:2021yvk,Bali:2014kia,DElia:2021tfb,Endrodi:2019zrl,Ding:2022tqn, Ding:2025pbu}. However, many of the commonly studied quantities in this context, such as chiral condensates, are primarily equilibrium diagnostics of QCD matter and do not map directly onto final-state observables in heavy-ion experiments. In contrast, conserved-charge fluctuations and correlations provide a more direct link between first-principles QCD thermodynamics and experiment, since they are computable on the lattice and can be related to event-by-event fluctuations of net baryon number ($\tB$), electric charge ($\tQ$), and strangeness ($\tS$)~\cite{HotQCD:2012fhj,Borsanyi:2011sw,Ding:2015ona,Bazavov:2017dus,Bollweg:2024epj,Clarke:2024ugt,Borsanyi:2025kiv,Adam:2026rnt,Clarke:2025gbj,Goswami:2026hit,Karsch:2010ck,Luo:2017faz,Rustamov:2022hdi,Pandav:2022xxx,Nonaka:2019fhk,Nonaka:2023xkg}. In magnetic backgrounds, first-principles studies of these observables are still relatively limited, making lattice-QCD calculations essential for establishing model-independent benchmarks.

Recent lattice-QCD calculations have begun to provide first-principles information on conserved-charge fluctuations and correlations in external magnetic fields. Initial studies with a larger-than-physical pion mass ($M_\pi\simeq 220~\rm MeV$) at a single lattice spacing~\cite{Ding:2021cwv} have recently been extended to physical pion mass~\cite{Ding:2023bft,Ding:2025jfz}.
These studies identified the baryon--electric-charge correlation $\chi^{\rm BQ}_{11}$ as a particularly sensitive response of QCD matter to a magnetic background, motivating its use as a QCD magnetometer~\cite{Ding:2023bft,Ding:2025jfz}. Along the pseudo-critical line, normalized ratios involving $\chi^{\rm BQ}_{11}$ exhibit pronounced magnetic-field enhancements, with ratios such as $\chi^{\rm BQ}_{11}/\chi^{\rm QS}_{11}$ and $\chi^{\rm BQ}_{11}/\chi^{\rm Q}_{2}$ among the most sensitive observables~\cite{Ding:2025jfz}. Related studies based on effective models and hadronic approaches have also investigated conserved-charge fluctuations and correlations in external magnetic fields, providing complementary perspectives on the magnetic-field dependence of charge-sector observables~\cite{Fu:2013ica,Fukushima:2016vix,Kadam:2019rzo,Chahal:2023khc,Mao:2024gox,Mao:2025toi,Mao:2026bfc,Samanta:2025mrq,Vovchenko:2024wbg}. The recent ALICE measurement of centrality-dependent conserved-charge correlations provides a first qualitative comparison to these lattice predictions for $\chi^{\rm BQ}_{11}/\chi^{\rm Q}_{2}$~\cite{ALICE:2025mkk}.

The same conserved-charge susceptibilities also determine the leading-order coefficients of the conserved-charge chemical potentials under strangeness neutrality and fixed charge-to-baryon ratio $r\equiv n_\tQ/n_\tB$. At small baryon chemical potential, these coefficients are defined through $\hat\mu_\tQ=q_1\hat\mu_\tB+O(\hat\mu_\tB^3)$ and $\hat\mu_\tS=s_1\hat\mu_\tB+O(\hat\mu_\tB^3)$, or equivalently $q_1\equiv(\mu_\tQ/\mu_\tB)_{\rm LO}$ and $s_1\equiv(\mu_\tS/\mu_\tB)_{\rm LO}$. In particular, $q_1$ shows a strong magnetic-field dependence, while $s_1$ responds more mildly~\cite{Ding:2023bft,Ding:2025nyh}. Since both coefficients depend on the isospin parameter $r$, they provide natural first-principles inputs for studying chemical-potential differences between collision systems with different isospin compositions.

The isobar partners $^{96}_{40}{\rm Zr}$ and $^{96}_{44}{\rm Ru}$ have the same mass number, $A=96$, but different charge-to-baryon ratios, $r_{\rm Zr}\simeq0.417$ and $r_{\rm Ru}\simeq0.458$. They therefore provide a controlled setting in which the isospin content can be varied while the mass number is kept fixed. Collisions of these systems were originally proposed to help disentangle the chiral magnetic effect from flow-driven and charge-conservation backgrounds~\cite{Voloshin:2010ut,Huang:2017azw,Deng:2018dut,STAR:2019bjg}. They have since also been used to explore nuclear-structure effects, baryon and electric-charge transport, and possible chiral-magnetic-effect signals in transport models~\cite{Li:2019kkh,Xu:2021vpn,STAR:2021mii,Pihan:2023dsb,Sun:2018idn,Kharzeev:2022hqz,Yuan:2024wpz}. For the present purpose, the difference between $r_{\rm Zr}$ and $r_{\rm Ru}$ directly leads to different values of $q_1(r)$ and $s_1(r)$ under strangeness neutrality, and thus provides a clean handle on isospin-driven chemical-potential splittings.

On the experimental side, recent Bayesian thermal analyses of STAR isobar data have extracted the chemical-potential differences $\Delta\mu_\tB$, $\Delta\mu_\tQ$, and $\Delta\mu_\tS$ between the Ru$+$Ru and Zr$+$Zr systems, with reduced systematic uncertainties obtained through double ratios of particle yields~\cite{Grefa:2026meq,STAR:2024lvy}. 
These studies found consistency in both sign and magnitude with lattice-QCD-based expectations at vanishing magnetic field, providing an important phenomenological benchmark. Despite these advances, a first-principles determination of isospin-driven chemical-potential splittings and their behavior in nonzero magnetic fields is still lacking. This motivates a dedicated lattice-QCD study of these quantities, based directly on conserved-charge susceptibilities at physical quark masses.

In this work, we present first-principles $(2+1)$-flavor lattice-QCD results for isospin-driven splittings of conserved-charge chemical potentials relevant for the Ru$+$Ru and Zr$+$Zr isobar systems, both at vanishing and nonzero magnetic fields. Under strangeness neutrality and fixed charge-to-baryon ratio, we outline a splitting framework that maps the isospin asymmetry encoded in $r_{\rm Zr}$ and $r_{\rm Ru}$ onto the experimentally accessible ratios $\Delta\mu_\tQ/\Delta\mu_\tB$, $\Delta\mu_\tS/\Delta\mu_\tB$, and $\Delta\mu_\tS/\Delta\mu_\tQ$, expressed as functions of $\mu_\tB(r_{\rm Ru})/\Delta\mu_\tB$. Using continuum-estimated lattice results for  $q_1$ and $s_1 $, we compare the $eB=0$ splitting ratios with the Bayesian thermal-analysis extractions from isobar data~\cite{Grefa:2026meq}, and extend the study to nonzero magnetic fields along $T_{\rm pc}(eB)$. Since the splitting ratios themselves show only weak magnetic-field dependence, we further examine isospin-driven differences of normalized chemical-potential ratios, in particular, $\mu_\tQ/\mu_\tB$.
To facilitate comparison with experiment, we also present hadron resonance gas (HRG) calculations and experimentally motivated proxy observables with kinematic cuts. 

The paper is organized as follows. Sec.~\ref{sec:isospin-driven-framework} introduces the strangeness-neutrality constraints relevant for heavy-ion collisions and outlines the framework used to study isospin-driven splittings of conserved-charge chemical potentials between Ru$+$Ru and Zr$+$Zr isobar systems. Sec.~\ref{sec:setup} describes the lattice setup and simulation parameters. Sec.~\ref{sec:results_q1s1} presents lattice results for the $r$--$eB$ dependence of the leading-order coefficients of the chemical-potential ratios $\mu_\tQ/\mu_\tB$ and $\mu_\tS/\mu_\tB$ along $T_{pc}(eB)$ under different isospin constraints. Sec.~\ref{sec:results_Dmu} reports lattice-QCD results for isospin-driven chemical-potential splitting ratios in the experimentally relevant regime in the presence of magnetic fields, together with HRG results and experimentally motivated proxy observables. We also discuss normalized ratio observables $R[\mathcal{O}]$ and their isospin-driven differences $\Delta R^{\rm Ru-Zr}[\mathcal{O}]$, as sensitive probes of magnetic-field effects in the two isobar systems. Finally, Sec.~\ref{sec:summary} summarizes our findings and presents our conclusions. The appendices collect supporting material for the main analysis: Appendix~\ref{app:NLO} estimates the size of next-to-leading-order corrections to the splitting-ratio framework, and Appendix~\ref{app:full_muBoverDmuB} extends splitting ratio results to a wider $\mu_\tB(r_{\rm Ru})/\Delta\mu_\tB$ window.

\section{Framework for Isospin-driven chemical-potential splittings}
\label{sec:isospin-driven-framework}

The QCD thermodynamic pressure encodes the bulk equilibrium properties of strongly interacting matter and their response to temperature ($T$), quark chemical potentials ($\hat{\mu}_f \equiv \mu_f / T$) for each flavor $f$, and magnetic field strength ($eB$). For a system of volume $V$, it is determined by the grand canonical partition function $\mathcal{Z}$ via
\beq
\hat{p} \equiv \frac{p}{T^4} = \frac{1}{VT^3} \ln \mathcal{Z}(T, V, \hat{\mu}_f, eB).
\eeq
The quark flavor chemical potentials are related to the conserved-charge basis of strong interactions through
\bea
\hmu_u &=& \frac{1}{3} \hmu_\tB + \frac{2}{3} \hmu_\tQ\,, \nn \\ 
\hmu_d &=& \frac{1}{3} \hmu_\tB - \frac{1}{3} \hmu_\tQ\,, \nn \\
\hmu_s &=& \frac{1}{3} \hmu_\tB - \frac{1}{3} \hmu_\tQ - \hmu_\tS\,,
\label{eq:mu_uds_BQS}
\eea
where $\hmu_{\rm B,Q,S}\equiv \mu_{\tB,\tQ,\tS}/T$ denote the chemical potentials conjugate to conserved charges: net baryon number $\tB$, electric charge $\tQ$, and strangeness $\tS$.

Fluctuations and correlations of conserved charges, $ \chi^{\tB \tQ \tS}_{ijk}(T,eB)$, defined as derivatives of the pressure with respect to chemical potentials, can be evaluated at vanishing chemical potentials, 
\bea
\label{eq:cc_suscp}
\quad \chi^{\tB \tQ \tS}_{ijk}(T,eB) = \frac{\partial^{i+j+k}}{\partial \hat{\mu}_{\tB}^{i} \partial \hat{\mu}_{\tQ}^j \partial  \hat{\mu}_{\tS}^k} \hat{p} \Bigg|_{\hat{\mu}_{\tB,\tQ,\tS}=0}\,,
\eea
where the charge-conjugation invariance dictates that only susceptibilities with an even sum of indices ($\{i,j,k\} \in \mathbb{Z}$ and $i + j + k \in 2\mathbb{Z}$) are nonzero. In this work, we focus on the second-order (leading-order) susceptibilities with $i+j+k=2$. Further details can be found in Refs.~\cite{DElia:2010abb,HotQCD:2012fhj,Petreczky:2012rq,Bazavov:2017dus,Petreczky:2012rq, Ding:2025nyh}.

In heavy-ion collision experiments, the colliding nuclei are initially net-strangeness neutral, and their valence quark content puts constraints on the conserved-charge sectors. Lattice-QCD studies consistently incorporate these physical constraints by imposing relevant strangeness neutrality and isospin asymmetry conditions on the conserved-charge chemical potentials \cite{Bazavov:2012vg, Bazavov:2014xya, Bollweg:2021vqf}:
\bea
 \label{eq:strange-neutral_isospin}
 \hat{n}^{\tS} = 0,\quad  {n}^{\tQ} /  {n}^{\tB} = r.
\eea
Here, $\hat{n}^X(T,eB,\hat{\mu}_{\rm B,Q,S})\equiv\hat{n}^X= \partial \hat{p}/\partial \hat{\mu}_X$ for $X\in \{\rm B,~Q,~S\}$ denotes the conserved-charge number densities in a hot magnetized medium. At sufficiently small $\hat{\mu}_\tB$, relevant to high-energy heavy-ion collisions, the analytic nature of thermodynamic observables around $\hat{\mu}_\tB=0$ allows Taylor expansion in $\hat{\mu}_\tB$. For the leading order in $\hat{\mu}_\tB$, $\hat{n}^X$ can be expressed in terms of second-order susceptibilities as detailed in our recent QCD equation of state work~\cite{Ding:2025nyh, Ding:2025qzh}.

To implement the strangeness neutrality constraint in QCD thermodynamics, we Taylor expand the electric charge and strangeness chemical potentials, $\hat{\mu}_{\{\tQ,\tS\}} \equiv \hat{\mu}_{\{\tQ,\tS\}}(T,eB,\hat{\mu}_{\tB})$ to the baryon sector:
\bea
\label{eq:muQ}
 \hat{\mu}_{\tQ} 
&=& q_1 (r,T,eB) \hat{\mu}_{\tB} + q_3 (r,T,eB) \hat{\mu}_{\tB}^3+ \mathcal{O}(\hat{\mu}_{\tB}^5)\,,\\
\label{eq:muS}
\hat{\mu}_{\tS} 
&=& s_1 (r,T,eB) \hat{\mu}_{\tB} + s_3 (r,T,eB) \hat{\mu}_{\tB}^3+ \mathcal{O}(\hat{\mu}_{\tB}^5).
\eea
The leading-order coefficients $q_1(r,T,eB)\equiv \left({\mu_\tQ}\big/{\mu_\tB} \right)_{\rm LO}$ and $s_1(r,T,eB)\equiv\left({\mu_\tS}\big/{\mu_\tB} \right)_{\rm LO}$ follow from enforcing the constraints in~\autoref{eq:strange-neutral_isospin} at leading order in $\hat{\mu}_\tB$~\cite{Bazavov:2012vg},
\bea
q_1 = \frac{r \left( \chi_{2}^{\tB}  \chi_{2}^{\tS} - \chi_{11}^{\tB \tS} \chi_{11}^{\tB \tS} \right) - \left(\chi_{11}^{\tB \tQ} \chi_{2}^{\tS} - \chi_{11}^{\tB \tS} \chi_{11}^{\tQ \tS} \right)}{\left(\chi_{2}^{\tQ } \chi_{2}^{\tS} - \chi_{11}^{\tQ \tS}  \chi_{11}^{\tQ \tS}\right) - r\left( \chi_{11}^{\tB \tQ} \chi_{2}^{\tS} -\chi_{11}^{\tB \tS} \chi_{11}^{\tQ \tS} \right)},
\label{eq:q1_def}\\
s_1 = - {\left( \chi_{11}^{\tB \tS} + q_1 \chi_{11}^{\tQ \tS}\right)}/ { \chi_{2}^{\tS}},
\label{eq:s1_def}
\eea
expressed as explicit functions of the second-order susceptibilities and the isospin parameter $r$.

The isospin parameter $r$ is fixed by the charge-to-baryon ratio of the colliding nuclei. Different collision systems therefore lead to different values of $q_1(r)$ and $s_1(r)$, and hence to system-dependent electric-charge and strangeness chemical potentials even at similar baryon density. Systems relevant for heavy-ion collisions include $r_{\rm Pb/Au}\simeq0.4$ for Pb+Pb and Au+Au collisions, $r_{\rm Zr}\simeq0.417$ for $^{96}_{40}$Zr, and $r_{\rm Ru}\simeq0.458$ for $^{96}_{44}$Ru. By contrast, systems such as $^{16}_{~8}$O and $^{20}_{10}$Ne correspond to the isospin-symmetric value $r_{\rm sym}=0.5$.

As discussed in Sec.~\ref{sec:intro}, the isobar collision systems $^{96}_{40}$Zr and $^{96}_{44}$Ru differ by $\sim10\%$ in their charge-to-baryon ratio, providing a controlled setting to study isospin-driven chemical-potential differences. 
In the following, all quantities are understood to be evaluated at fixed $T$ and $eB$; hats are therefore omitted in ratios of chemical potentials. We define the Zr--Ru splitting convention by
\beq
\Delta\hmu_X \equiv \hmu_X(r_{\rm Zr}) - \hmu_X(r_{\rm Ru}), \quad X\in \{\rm B,~Q,~S\}.
\label{eq:DmuX}
\eeq

Using the leading-order relations $\hmu_\tQ=q_1(r)\hmu_\tB$ and $\hmu_\tS=s_1(r)\hmu_\tB$, with the dependence on $T$ and $eB$ suppressed for notational simplicity, one obtains\footnote{Note that the next-to-leading-order corrections, briefly outlined in App.~\ref{app:NLO}, are suppressed by $\mathcal{O}(10^{-3})$ relative to leading-order ratios near the isobar freeze-out point. Thus we only consider the leading-order expansions thereafter.}
\bea
\Delta\hmu_\tQ \equiv q_{1}(r_{\rm Zr}) \hmu_\tB(r_{\rm Zr})  -q_{1}(r_{\rm Ru}) \hmu_\tB(r_{\rm Ru})  \,,\\
\Delta\hmu_\tS \equiv s_{1}(r_{\rm Zr}) \hmu_\tB(r_{\rm Zr}) - s_{1}(r_{\rm Ru}) \hmu_\tB(r_{\rm Ru}).
\eea
Substituting Eq.~\eqref{eq:DmuX} for the baryon chemical potential gives
\bea
\Delta\hmu_\tQ \equiv  q_{1}(r_{\rm Zr}) \left(\hmu_\tB(r_{\rm Ru}) + \Delta\hmu_\tB \right) - q_{1}(r_{\rm Ru}) \hmu_\tB(r_{\rm Ru})  \,,\\
\Delta\hmu_\tS \equiv s_{1}(r_{\rm Zr}) \left(\hmu_\tB(r_{\rm Ru}) + \Delta\hmu_\tB \right) - s_{1}(r_{\rm Ru}) \hmu_\tB(r_{\rm Ru})  \,.
\eea

This yields the ratios of chemical-potential splittings $\Delta\hmu_{\{\tQ,\tS\}}$ over $\Delta\hmu_\tB$ as functions of $\mu_\tB(r_{\rm Ru})/\Delta\mu_\tB$: 
\bea
\frac{\Delta \mu_\tQ}{\Delta \mu_\tB } \equiv q_{1}(r_{\rm Zr})+\Big[q_{1}(r_{\rm Zr}) -  q_{1}(r_{\rm Ru}) \Big] \frac{\mu_\tB(r_{\rm Ru})}{\Delta \mu_\tB },
\label{eq:DmuQ_DmuB}
\\ 
\frac{\Delta \mu_\tS}{\Delta \mu_\tB} \equiv s_1(r_{\rm Zr}) + \Big[s_{1}(r_{\rm Zr})-s_{1}(r_{\rm Ru}) \Big]\frac{ \mu_\tB(r_{\rm Ru})}{\Delta \mu_\tB} 
\label{eq:DmuS_DmuB}.
\eea

The corresponding strangeness-to-electric-charge splitting ratio is then
\bea
\frac{\Delta \mu_\tS}{ \Delta \mu_\tQ} &\equiv& \frac{\Delta \mu_\tS}{\Delta \mu_\tB } \Big/ \frac{\Delta \mu_\tQ}{\Delta \mu_\tB }  \nn \\ 
&=& \frac{s_1(r_{\rm Zr})+\big(s_1(r_{\rm Zr})  - s_1(r_{\rm Ru})\big)\frac{\mu_\tB(r_{\rm Ru})}{\Delta \mu_\tB} }{q_1(r_{\rm Zr})+\big(q_1(r_{\rm Zr})  - q_1(r_{\rm Ru})\big)\frac{\mu_\tB(r_{\rm Ru})}{\Delta \mu_\tB} }.
\label{eq:DmuS_DmuQ}
\eea

 The chemical-potential splitting ratios in Eqs.~\eqref{eq:DmuQ_DmuB}--\eqref{eq:DmuS_DmuQ} are determined,
on the theory side, by the first-principles lattice-QCD inputs
$q_1(r,T,eB)$ and $s_1(r,T,eB)$, together with the freeze-out ratio
$\mu_B(r_{\rm Ru})/\Delta\mu_B$ extracted from the Bayesian thermal
analysis of isobar-collision data~\cite{Grefa:2026meq}.  This construction
therefore provides lattice-QCD benchmarks for the isospin-driven
chemical-potential splittings between the Ru+Ru and Zr+Zr systems.
The comparison with Ref.~\cite{Grefa:2026meq} should be understood as a
comparison with thermal-analysis-extracted chemical potentials, rather
than with directly measured observables, and is subject to the associated freeze-out, modeling, and collision-dynamical uncertainties~\cite{Andronic:2005yp,Cleymans:2005xv,Andronic:2017pug,Bazavov:2012vg,Karsch:2015zna}.

\begin{figure*}[htbp]
    \centering
    \includegraphics[width=0.45\linewidth]{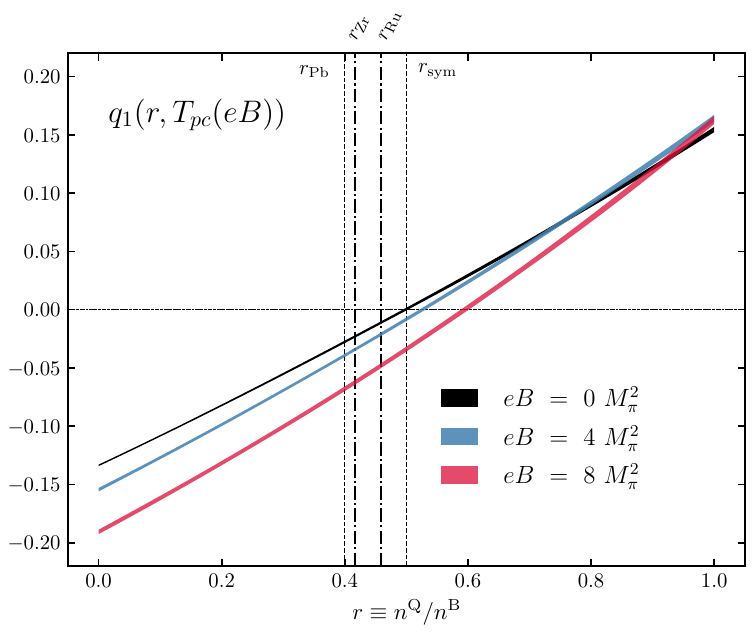}
    \includegraphics[width=0.45\linewidth]{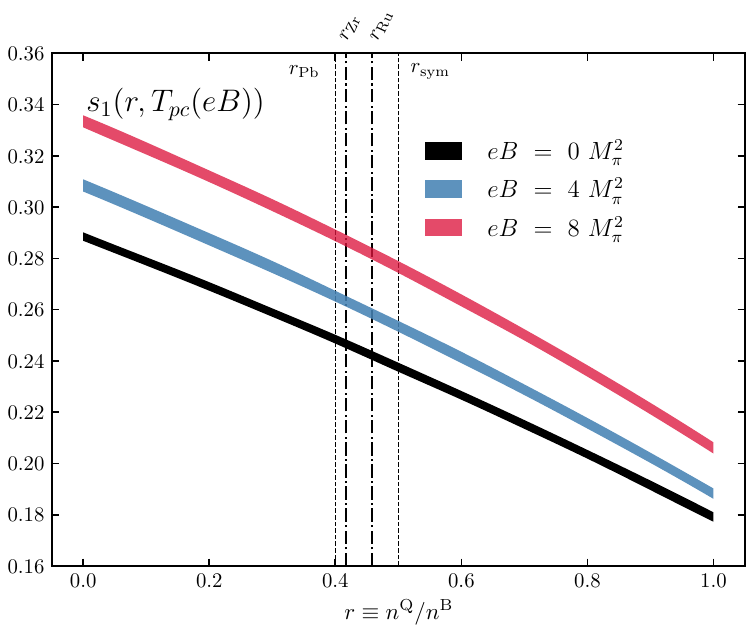}
    
    \caption{Isospin parameter $r$-dependence of the leading-order coefficients $q_1(r,T,eB) \equiv (\mu_\tQ/\mu_\tB)_{\rm LO}$ (left), and $s_1(r,T,eB) \equiv (\mu_\tS/\mu_\tB)_{\rm LO}$ (right) along the pseudo-critical line $T_{pc}(eB)$. Colored bands represent lattice-QCD continuum estimates at various magnetic field strengths. Broken vertical lines mark the isospin parameters relevant to heavy-ion collision systems, including Zr$+$Zr and Ru$+$Ru isobar collisions.}
    \label{fig:q1s1_PbZrRu_Tpc-eB_r}
\end{figure*}

\section{Lattice setup}
\label{sec:setup}

The QCD partition function is given by the functional integral
\bea
\mathcal{Z} &=&  \int \mathcal{D} U~ \prod_{f=\{u,d,s\}}\left[{\rm det} {M_f}(U,B,q_f,m_f,\mu_f) \right]^{1/4} \nn \\
&&\qquad \qquad \qquad \qquad \times ~ e^{-S_{g}(\beta, U)}\,,
\eea
where $M_f(U,B,q_f,m_f,\mu_f)$ denotes the fermion matrix for flavor $f$ in the presence of a background magnetic field $B$, quark electric charge $q_f$, mass $m_f$ (with degenerate light quarks $m_u=m_d$), and chemical potential $\mu_f$. The gauge action is parameterized by $\beta = 6/g^2$. We perform $(2+1)$-flavor lattice-QCD simulations employing the highly improved staggered quark (HISQ) action \cite{Follana:2006rc} together with the tree-level improved Symanzik gauge action. This setup follows the framework extensively used by the HotQCD collaboration \cite{Bazavov:2019www,Bollweg:2021vqf,Bollweg:2022rps,Bollweg:2024epj}.

A uniform magnetic field $\vec{B}=(0,0,B)$ is introduced along the $z$ direction through U(1) link variables implemented in Landau gauge \cite{DElia:2010abb,Bali:2011qj,Al-Hashimi:2008quu}, which multiply the SU(3) gauge links.  Magnetic-field quantization follows from the finite lattice geometry and periodic boundary conditions,
\beq
eB = \frac{6\pi N_b}{N_x N_y}a^{-2}, \quad N_b \in \mathbb{Z},
\eeq
where $e$ is the elementary electric charge, and $N_b$ counts magnetic flux quanta through the $x$–$y$ plane for lattice points $N_x,N_y$ with spacing $a$. The quark electric charges satisfy $q_d=q_s=-q_u/2=-e/3$ and to ensure flux quantization for all flavors simultaneously, we adopt the greatest common divisor of the electric charges, $\left|q_d\right|=\left|q_s\right|=-\left|q_u\right|/2=e/3$. Furthermore, the periodicity of U(1) links implies a constraint, $0\le N_b < N_xN_y/4$, on magnetic flux and in this work, we restrict to $N_b \le 6$, corresponding to magnetic fields up to $eB \lesssim 8 M_\pi^2$. For these values, discretization effects associated with the magnetic field remain controlled, as $N_b/N_\sigma^2 \ll 1$. Further details about incorporating magnetic fields with the HISQ action can be found in Refs.~\cite {Ding:2020hxw, Ding:2021cwv}.

We employ spatially symmetric lattices with $N_\sigma \equiv N_x=N_y=N_z$ and fixed aspect ratio $N_\sigma/N_\tau=4$, primarily using $32^3\times 8$ and $48^3\times 12$ lattices. The lattice-QCD results are continuum estimated based on these $N_\tau=8$ and $12$ lattices; see Refs.~\cite{Ding:2023bft,Ding:2025jfz,Ding:2025nyh} for further details. The strange quark mass is tuned to its physical value, with degenerate light quarks satisfying $m_u=m_d=m_s/27$, corresponding to a pseudo-Goldstone pion mass $M_\pi \simeq 135~{\rm MeV}$ at vanishing magnetic field. Scale setting follows Refs.~\cite{Bollweg:2021vqf,Bazavov:2019www}. For $N_b=0$, we adopt lattice results from Ref.~\cite{Bollweg:2021vqf}.

\section{Leading-order coefficients under strangeness neutrality and fixed charge-to-baryon ratio}
\label{sec:results_q1s1}

\begin{figure}[htbp]
    \centering
    \includegraphics[width=0.9\linewidth]{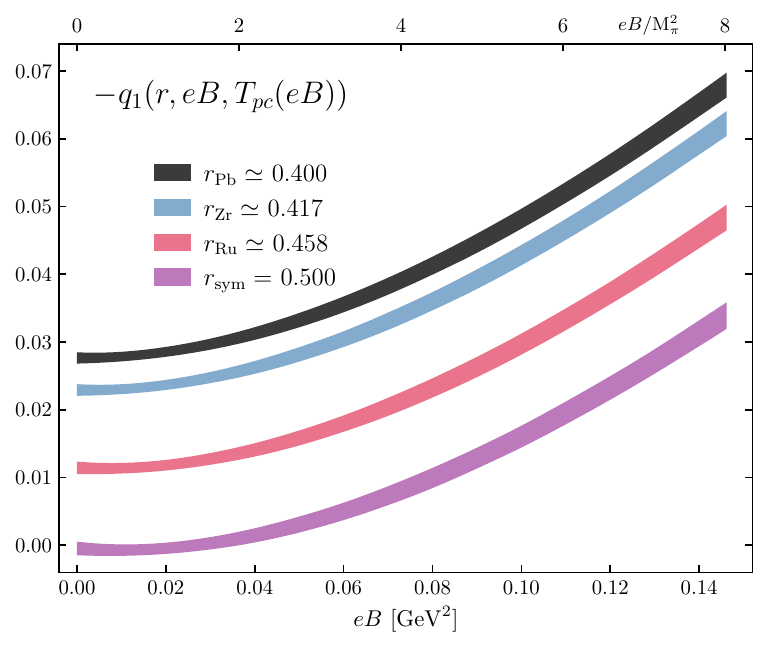}
    \includegraphics[width=0.9\linewidth]{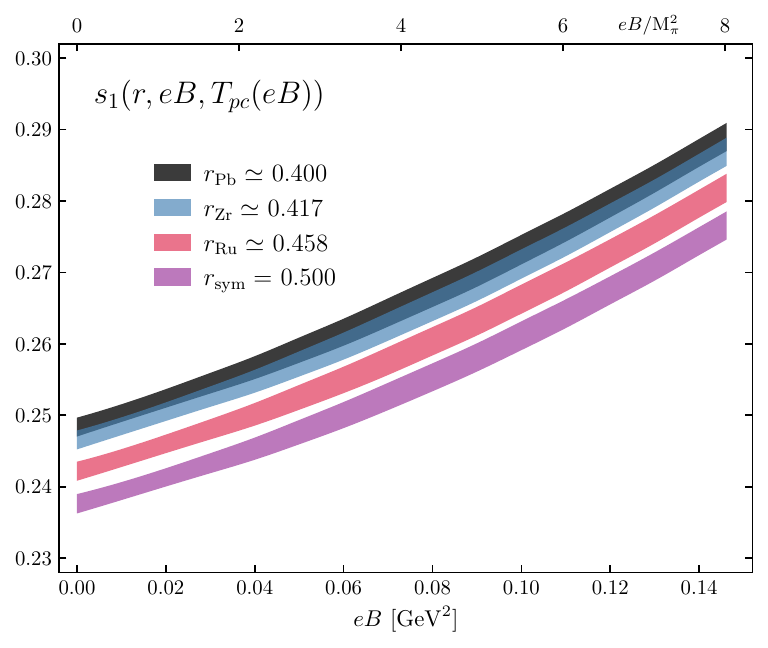}

    \includegraphics[width=0.9\linewidth]{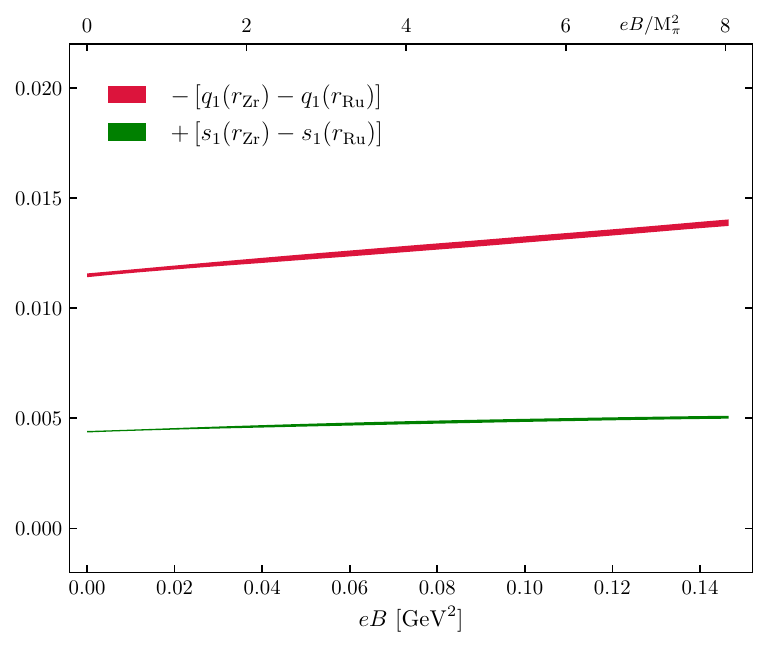}

    \caption{Magnetic field $eB$-dependence of the leading-order coefficients $-q_1$ (top) and $s_1$ (middle) along the pseudo-critical line $T_{pc}(eB)$ for the fixed isospin parameters $r_{\rm Pb}\simeq0.4$, $r_{\rm Zr}\simeq0.417$, $r_{\rm Ru}\simeq0.458$, and $r_{\rm sym}=0.5$. The bottom panel shows the isospin-driven difference coefficients $-[q_1(r_{\rm Zr})-q_1(r_{\rm Ru})]$ and $s_1(r_{\rm Zr})-s_1(r_{\rm Ru})$ for the isobar systems. The colored bands represent lattice-QCD continuum estimates.}    \label{fig:q1s1_PbZrRu_Tpc-eB}
\end{figure}

Within the framework described in the previous section, the chemical-potential splittings are primarily controlled by the sign and magnitude of the leading-order electric-charge-to-baryon chemical-potential ratio, $q_1 \equiv (\mu_\tQ/\mu_\tB)_{\rm LO}$, and the corresponding strangeness-to-baryon chemical-potential ratio, $s_1 \equiv (\mu_\tS/\mu_\tB)_{\rm LO}$, under strangeness neutrality. In this section, we present lattice-QCD results for these coefficients, $q_1(r,eB, T_{pc}(eB))$ and $s_1(r,eB, T_{pc}(eB))$, and analyze their dependence on the isospin parameter $r$ and magnetic field strength $eB$ along the pseudo-critical line $T_{pc}(eB)$.

\subsection{$r$-dependence of $q_1$ and $s_1$ along $T_{pc}(eB)$}
\label{sec:vsr_q1s1}

We begin by examining the isospin-parameter dependence of $q_1$ and $s_1$. \autoref{fig:q1s1_PbZrRu_Tpc-eB_r} presents the continuum estimates for $q_1(r)$ (left) and $s_1(r)$ (right) as functions of the isospin parameter $r$ along $T_{pc}(eB)$\footnote{The pseudo-critical line $T_{pc}(eB)$ of the QCD chiral crossover in nonzero magnetic fields was determined from the peak location of the total chiral susceptibility in Ref.~\cite{Ding:2025jfz}. Notably, $T_{pc}$ remains nearly unchanged across the considered $eB$ range: $T_{pc}\simeq155.8(9)$, $155.3(4)$, and $155.1(4)~{\rm MeV}$ at $eB=0$, $4M_\pi^2$, and $8M_\pi^2$, respectively.} for $eB = 0$, $4M_\pi^2$, and $8M_\pi^2$. In principle, under strangeness neutrality, $r$ can take values in the full range $0\le r\le 1$.  For heavy-ion
applications, however, the mildly isospin-asymmetric region
$r\simeq 0.4$--$0.5$ is most relevant. The broken vertical lines mark the specific values $r_{\rm Pb} \simeq 0.4,\, r_{\rm Zr} \simeq 0.417,\,$ and $r_{\rm Ru} \simeq 0.458$, relevant to Pb$+$Pb, Zr$+$Zr, and Ru$+$Ru collision systems, respectively, together with the symmetric case $r_{\rm sym} = 0.5$.

In the left panel, we observe that the leading-order coefficient $q_1(r)$ exhibits a sign change controlled by isospin parameter $r$,  which specifies the charge-to-baryon ratio.
At vanishing magnetic
field, $q_1(r)$ is negative for $r<0.5$ and positive for $r>0.5$, while it
vanishes at the isospin-symmetric point $r=r_{\rm sym}=0.5$.  This behavior
reflects the role of $\mu_\tQ$ in enforcing the imposed charge-to-baryon ratio:
for systems with a smaller electric-charge fraction than the symmetric case,
a negative electric-charge chemical potential is required. A nonzero magnetic field modifies this pattern in a systematic way. For $r < 0.5$, including the heavy-ion-collision--relevant cases $r \in \{ r_{\rm Pb}, r_{\rm Zr}, r_{\rm Ru} \}$, $q_1(r)$ becomes more negative with increasing magnetic-field strength. Notably, an apparent isospin hierarchy emerges in $q_1(r, eB, T_{pc}(eB))$,
\beq
|q_1(r_{\rm Pb})|>|q_1(r_{\rm Zr})| >|q_1(r_{\rm Ru})|\,,
\eeq 
which persists over the magnetic-field range considered. The magnetic field also shifts the zero crossing of $q_1(r)$ to values larger than $r=0.5$.  This can be understood as a consequence of magnetic-field--induced enhancements in the population of positively charged baryons through the increased degeneracy of their lowest Landau levels, thereby favoring charged states over neutral ones. To maintain the fixed charge-to-baryon ratio, $\mu_\tQ$ has to therefore adjust and become increasingly negative.

In the strangeness sector (right panel), the effects of strangeness neutrality are encoded in $s_1(r)$, which arises primarily from baryon-strangeness correlations and receives an indirect $r$-dependence through the electric-charge constraint. Unlike $q_1(r)$, $s_1(r)$ remains positive over the full range of $r$, reflecting the fact that 
strange quarks carry negative strangeness and that a positive $\mu_\tS$ is required to maintain strangeness neutrality. Magnetic fields induce a mild enhancement of $s_1(r)$ across all values of $r$, including the heavy-ion--collision--relevant cases, qualitatively similar to the $eB$-enhancement observed for $|q_1(r)|$, though substantially weaker in magnitude.

From the perspective of the isospin-driven chemical potential splitting framework, the intercept coefficients in Eqs.~\eqref{eq:DmuQ_DmuB}--\eqref{eq:DmuS_DmuQ} satisfy $q_1(r_{\rm Zr}) < 0$ and $s_1(r_{\rm Zr}) > 0$. Since $r_{\rm Ru} > r_{\rm Zr}$, the hierarchy discussed above implies $q_1(r_{\rm Ru}) > q_1(r_{\rm Zr})$, corresponding to a less negative value of $q_1$ at $r_{\rm Ru}$, alongside $s_1(r_{\rm Ru}) < s_1(r_{\rm Zr})$. 
These ordering relations, combined with the framework of Sec.~\ref{sec:isospin-driven-framework}, will be used in Sec.~\ref{sec:results_Dmu} to determine the sign structure and relative magnitude of the chemical-potential splittings.

\subsection{$eB$-dependence at fixed isospin parameters}
\label{sec:q1s1_RuZr}

Having established the isospin hierarchy of $q_1(r)$ and $s_1(r)$ across the relevant $r$-range, we now examine their magnetic-field dependence at fixed isospin parameters relevant to heavy-ion collisions, $r \in \{ r_{\rm Pb}, r_{\rm Zr}, r_{\rm Ru},r_{\rm sym} \}$. This allows us to directly compare the Ru$+$Ru and Zr$+$Zr isobar systems as the magnetic field is varied.

\autoref{fig:q1s1_PbZrRu_Tpc-eB} presents $-q_1(r,eB,T_{pc}(eB))$ (top) and $s_1(r,eB,T_{pc}(eB))$ (middle) as functions of the magnetic field strength $eB$ along the pseudo-critical line $T_{pc}(eB)$ for the above-mentioned systems.  In the top panel, continuum estimates for $-q_1$ show a pronounced magnetic-field sensitivity across all considered fixed-$r$ scenarios, exhibiting a monotonic enhancement with $eB$. Furthermore, the isospin hierarchy established in the previous subsection is clearly visible in this fixed-$r$ representation. 
In particular, the separations between the continuum bands reflect the different isospin parameters of the corresponding systems and are maintained as the magnetic-field strength increases.

By contrast, $s_1$ in the middle panel exhibits substantially weaker sensitivity in this $r$--$eB$ regime. Although $s_1$ grows monotonically with $eB$ for all considered systems, the enhancements are significantly suppressed compared to its electric-charge counterpart. The separations between the corresponding continuum bands remain relatively small, indicating a comparatively mild isospin sensitivity.

To quantify the isobar contrast directly, we define the isospin-driven difference coefficients as $\mathcal{O}(r_{\rm Zr},T,eB) - \mathcal{O}(r_{\rm Ru},T,eB)$ for $\mathcal{O} \in \{ q_1, s_1 \}$. In the bottom panel of \autoref{fig:q1s1_PbZrRu_Tpc-eB}, we present lattice-QCD continuum estimates for $-\left[q_1(r_{\rm Zr})-q_1(r_{\rm Ru})\right]$ and $\left[s_1(r_{\rm Zr})-s_1(r_{\rm Ru})\right]$ along the pseudo-critical line. Consistent with the isospin hierarchy discussed above, $\left[q_1(r_{\rm Zr})-q_1(r_{\rm Ru})\right]$ is negative and its magnitude grows mildly with the magnetic field strength. In contrast, the corresponding strangeness-sector difference $\left[s_1(r_{\rm Zr})-s_1(r_{\rm Ru})\right]$ is positive, while its absolute magnitude is roughly a factor of two smaller than 
$\left|q_1(r_{\rm Zr})-q_1(r_{\rm Ru})\right|$, and it exhibits negligible dependence on $eB$. The small uncertainties of these difference coefficients arise from the correlated determination of $q_1$ and $s_1$ at the two nearby isospin values. Since the two coefficients are evaluated on the same gauge ensembles and have similar $r$-dependence, their statistical fluctuations are strongly correlated. We therefore form the differences within each block-bootstrap sample, so that common fluctuations largely cancel in $q_1(r_{\rm Zr})-q_1(r_{\rm Ru})$ and $s_1(r_{\rm Zr})-s_1(r_{\rm Ru})$, leading to reduced uncertainties compared with the individual coefficients.

From Eqs.~(\ref{eq:DmuQ_DmuB}) and~(\ref{eq:DmuS_DmuB}), the sign and magnitude of the $\Delta\mu_\tQ$ and $\Delta\mu_\tS$ splittings are governed by the leading-order coefficients $q_1(r_{\rm Zr}),~s_1(r_{\rm Zr})$ and their isospin-driven differences. The coefficients themselves set the baseline intercept, while the isospin-driven differences control the slope, amplified by the factor $\mu_\tB(r_{\rm Ru})/\Delta\mu_\tB$. Since $q_1(r_{\rm Zr}) < 0$ and $\left[q_1(r_{\rm Zr})-q_1(r_{\rm Ru})\right] < 0$, both terms reinforce each other, yielding $\Delta\mu_\tQ<0$ for positive $\mu_\tB(r_{\rm Ru})/\Delta\mu_\tB$. Since magnetic fields significantly enhance $|q_1(r_{\rm Zr})|$, whereas the isospin-driven difference changes only mildly, the dominant magnetic-field dependence of $\Delta\mu_\tQ/\Delta\mu_\tB$ is expected to arise from the intercept term. Analogously, $s_1(r_{\rm Zr})>0$ and the much smaller strangeness-sector isospin difference imply $\Delta\mu_\tS>0$, with both magnetic-field and isospin sensitivities expected to be weaker than in the electric-charge sector.


\section{Isospin-driven splitting of conserved charge chemical potentials}
\label{sec:results_Dmu}

Having established the isospin and magnetic-field dependence of the leading-order coefficients $q_1$ and $s_1$ in Sec.~\ref{sec:results_q1s1}, we now combine our lattice-QCD results for the isobar-relevant coefficients and their respective isospin-driven differences with the chemical-potential splitting framework relations introduced in Sec.~\ref{sec:isospin-driven-framework}. In this section, we present our lattice-QCD results for the isospin-driven splittings of conserved-charge chemical potentials relevant in Zr$+$Zr and Ru$+$Ru heavy-ion collisions, expressed through the ratios
\beq 
{\Delta\mu_\tQ}\big/{\Delta\mu_\tB}, ~~ {\Delta\mu_\tS}\big/{\Delta\mu_\tB}, \text{and}~~ {\Delta\mu_\tS}\big/{\Delta\mu_\tQ}, 
\eeq 
as defined in Eqs.~(\ref{eq:DmuQ_DmuB}), (\ref{eq:DmuS_DmuB}), and~(\ref{eq:DmuS_DmuQ}), respectively. These ratios are functions of
$\mu_B(r_{\rm Ru})/\Delta\mu_B$, with their intercepts and slopes fixed
by $q_1(r_{\rm Zr})$, $s_1(r_{\rm Zr})$, and the isospin-driven
differences $q_1(r_{\rm Zr})-q_1(r_{\rm Ru})$ and
$s_1(r_{\rm Zr})-s_1(r_{\rm Ru})$. The lattice-QCD calculation therefore
determines the functional dependence of the splitting ratios, while the
Bayesian thermal-analysis extraction from isobar-collision hadron yields
identifies the phenomenologically relevant range of
$\mu_B(r_{\rm Ru})/\Delta\mu_B$. At vanishing magnetic field, the
corresponding comparison is made with the 0--10\% centrality results of
Ref.~\cite{Grefa:2026meq}.

\begin{figure}[!htbp]
    \centering
    \includegraphics[width=0.9\linewidth]{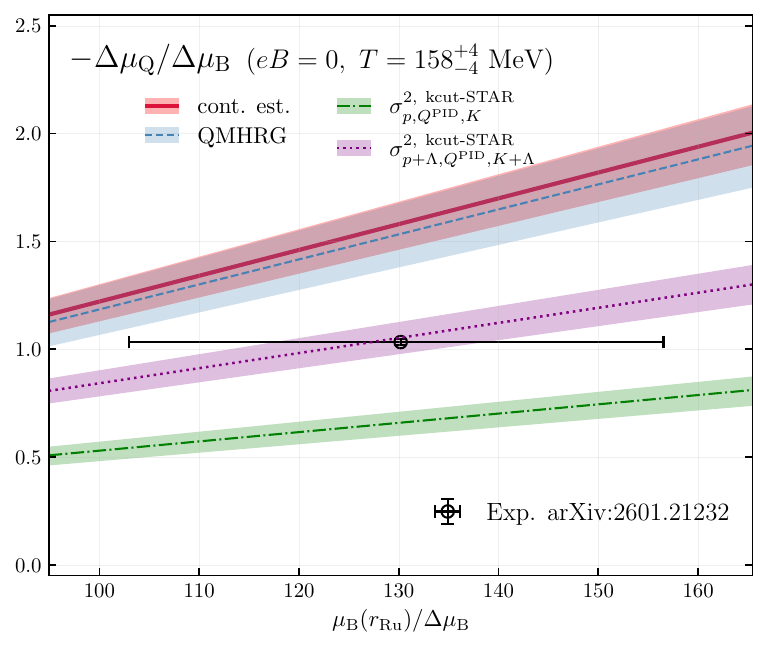}
    \includegraphics[width=0.9\linewidth]{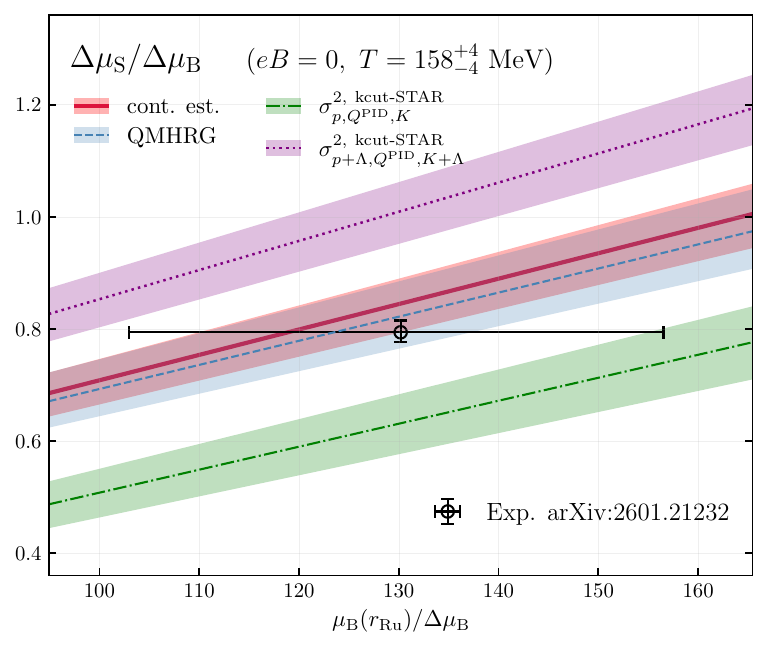}
    \includegraphics[width=0.9\linewidth]{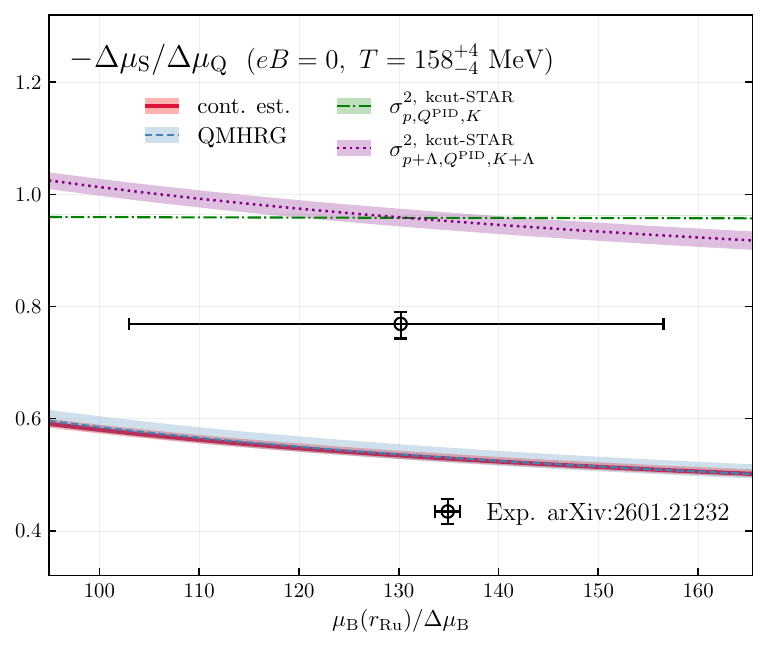}
    
    \caption{Isospin-driven conserved charge splitting ratios, ${-\Delta \mu_\tQ}/{ \Delta \mu_\tB} $, ${\Delta \mu_\tS}/{ \Delta \mu_\tB}$ and ${-\Delta \mu_\tS}/{ \Delta \mu_\tQ} $ from top to bottom, near the isobar freeze-out line $~T=158^{+4}_{-4}~{\rm MeV}$, denoted by shaded bands around central value. The red band represents the lattice-QCD continuum estimates, and the black circles denote Bayesian thermal-analysis extracted ratios and freeze-out region. Colored broken lines denote results within the HRG framework: dashed for full QMHRG, while the dash-dotted and dotted lines for STAR-cut proxy observables constructed from
$\pi,K,p$ and $\pi,K,p,\Lambda$ species, respectively. The STAR-motivated cuts are $|y|<0.5$ and $0.2<p_T<2.5~{\rm GeV}/c$ for $\pi, K$, $0.45<p_T<2.5~{\rm GeV}/c$ for $p$, and $0.4<p_T<1.6~{\rm GeV}/c$ for $\Lambda$.}
    \label{fig:RdmuBQS_Zr-Ru_T158-eB0}
\end{figure}

\autoref{fig:RdmuBQS_Zr-Ru_T158-eB0} presents lattice-QCD continuum estimates and HRG results for the isospin-driven conserved-charge splitting ratios $-{\Delta \mu_\tQ}/{ \Delta \mu_\tB}$ (top), ${\Delta \mu_\tS}/{ \Delta \mu_\tB}$ (middle), and $-{\Delta \mu_\tS}/{ \Delta \mu_\tQ}$ (bottom) in the absence of magnetic fields ($eB=0$). The lattice-QCD continuum estimates are shown as red bands at the isobar freeze-out temperature $T=158^{+4}_{-4}~{\rm MeV}$, plotted as functions of $\mu_\tB(r_{\rm Ru})/\Delta\mu_\tB$. The lighter shaded band spans the freeze-out temperature uncertainty. The displayed range of $\mu_\tB(r_{\rm Ru})/\Delta\mu_\tB$ is chosen to cover the 
phenomenologically relevant region identified by the Bayesian
thermal-analysis extraction.  
The black circles denote the
corresponding extracted values based on STAR isobar-collision
hadron-yield data~\cite{STAR:2024lvy,Grefa:2026meq}, with
$\mu_\tB(r_{\rm Ru})=20.31^{+3.8}_{-4.0}~{\rm MeV}$ and
$\Delta\mu_\tB=0.156^{+0.011}_{-0.012}~{\rm MeV}$ at $T=158.0^{+3.9}_{-3.9}~\rm MeV$. For visual clarity, the electric-charge splitting ratio and the
strangeness-to-electric-charge splitting ratio are shown with an
overall minus sign, since the lattice-QCD results give
$\Delta\mu_\tQ/\Delta\mu_\tB<0$ and
$\Delta\mu_\tS/\Delta\mu_\tQ<0$ throughout the relevant window. This
sign structure follows directly from the isospin hierarchy
$q_1(r_{\rm Zr})<q_1(r_{\rm Ru})<0$, together with the positive
strangeness-sector splitting discussed in Sec.~\ref{sec:results_q1s1}.

Beyond the overall signs, the dependence on
$\mu_\tB(r_{\rm Ru})/\Delta\mu_\tB$ is governed by the
isospin-driven difference terms in
Eqs.~(\ref{eq:DmuQ_DmuB}) and~(\ref{eq:DmuS_DmuB}). In the
electric-charge sector, $q_1(r_{\rm Zr})-q_1(r_{\rm Ru})<0$, so
$\Delta\mu_\tQ/\Delta\mu_\tB$ becomes increasingly negative as
$\mu_\tB(r_{\rm Ru})/\Delta\mu_\tB$ increases, or equivalently the
plotted quantity $-\Delta\mu_\tQ/\Delta\mu_\tB$ increases. In the
strangeness sector, the lattice results give
$s_1(r_{\rm Zr})-s_1(r_{\rm Ru})>0$ in the relevant range, leading to a
monotonic increase of $\Delta\mu_\tS/\Delta\mu_\tB$. Since the
electric-charge splitting grows more rapidly in magnitude than the
strangeness splitting, the magnitude of the strangeness-to-electric-charge
splitting ratio, $|\Delta\mu_\tS/\Delta\mu_\tQ|$, decreases mildly with
$\mu_\tB(r_{\rm Ru})/\Delta\mu_\tB$. Because the isospin-driven differences are so precisely determined (Sec.~\ref{sec:results_q1s1}), these lattice-QCD splitting-ratio bands remain statistically well constrained even in the large $\mu_\tB(r_{\rm Ru})/\Delta\mu_\tB$ experimental window.

Quantitatively, the Bayesian thermal-analysis extraction gives
$\Delta \mu_\tQ/\Delta \mu_\tB=-1.033^{+0.013}_{-0.012}$,
$\Delta \mu_\tS/\Delta \mu_\tB= 0.794^{+0.021}_{-0.017}$, and
$\Delta \mu_\tS/\Delta \mu_\tQ= -0.769^{+0.021}_{-0.026}$ at the
extracted values of $\mu_\tB(r_{\rm Ru})$, $\Delta\mu_\tB$, and $T$.\footnote{For the electric-charge and
strangeness sectors, Ref.~\cite{Grefa:2026meq} reports the inverse
ratios $\Delta\mu_\tB/\Delta\mu_\tQ=-0.968^{+0.011}_{-0.012}$ and
$\Delta\mu_\tB/\Delta\mu_\tS=1.259^{+0.028}_{-0.032}$. We have converted them to
$\Delta\mu_\tQ/\Delta\mu_\tB$ and $\Delta\mu_\tS/\Delta\mu_\tB$ using
standard error propagation.} Among the three ratios,
$\Delta\mu_\tS/\Delta\mu_\tB$ shows the closest agreement with the
lattice-QCD continuum estimate. The electric-charge splitting ratio
$\Delta\mu_\tQ/\Delta\mu_\tB$ has the correct sign and comparable
magnitude, but the lattice-QCD estimate is larger in magnitude than the
Bayesian-extracted value in the relevant window. Consequently, the
strangeness-to-electric-charge splitting ratio
$\Delta\mu_\tS/\Delta\mu_\tQ$ is slightly smaller in magnitude than the
Bayesian-extracted value. These splitting ratio comparisons are particularly sensitive to
the extracted value of $\mu_\tB(r_{\rm Ru})/\Delta\mu_\tB$, and hence to
the separate determinations of $\mu_\tB(r_{\rm Ru})$ and
$\Delta\mu_\tB$.

The HRG model provides a useful hadronic reference for interpreting these splitting ratios. In analogy with the lattice-QCD construction, the same relations in Eqs.~(\ref{eq:DmuQ_DmuB})--(\ref{eq:DmuS_DmuQ}) can be applied using conserved-charge susceptibilities obtained from the HRG thermodynamic pressure. The full QMHRG results (blue dashed lines), including the shaded bands associated with the freeze-out temperature uncertainty, follow the overall behavior of the lattice-QCD continuum estimates and remain compatible
with them in the phenomenologically relevant window. To make closer
contact with experimentally accessible quantities, we also construct
proxy observables for conserved charges within the PDGHRG framework~\cite{ParticleDataGroup:2020ssz}.
Following the particle content used in the Bayesian thermal analysis of
Ref.~\cite{Grefa:2026meq}, we use identified particles to approximate
$\text{net-B}\equiv \tilde{p}$,
$\text{net-Q}^{\rm PID}\equiv \tilde{\pi}^{+}+\tilde{K}^{+}+\tilde{p}$,
and $\text{net-S}\equiv \tilde{K}^{+}$. We further include the
reconstructed $\tilde{\Lambda}$ hyperon to improve the description of the
$\chi^{\rm BS}_{11}$ sector. For the detector-level proxies, we impose
STAR-motivated kinematic cuts on $\pi$, $K$, $p$, and $\Lambda$, following Refs.~\cite{Karsch:2015zna, Bellwied:2019pxh}, as
specified in the caption of Fig.~\ref{fig:RdmuBQS_Zr-Ru_T158-eB0}. Since the $\Lambda$ carries strangeness $S=-1$, it enters the net-strangeness proxy with the opposite sign to the kaon, i.e. $\text{net-S}\to\tilde{K}^{+}-\tilde{\Lambda}$; for simplicity, we set the subscript $K+\Lambda$ in the corresponding STAR-cut proxy $\sigma^{2,\rm kcut\text{-}STAR}_{p+\Lambda,\,\tQ^{\rm PID},\,K+\Lambda}$. The relevant chemical-potential observables are then constructed from these proxy fluctuations.

The proxy observables with kinematic cuts lead to visible deviations
from the full QMHRG and lattice-QCD results. In the electric-charge
sector, they suppress $-\Delta\mu_\tQ/\Delta\mu_\tB$, with the result
remaining below or close to unity depending on whether the
$\Lambda$-extended proxy set is used. The Bayesian-extracted value lies
between the lattice-QCD continuum estimate and the $\pi,K,p$ cut-proxy result,
suggesting that acceptance effects and the proxy definition of net
electric charge can have a sizable impact. In the strangeness sector,
$\Delta\mu_\tS/\Delta\mu_\tB$ is less strongly suppressed by the cuts.
The standard $\pi,K,p$ proxy, which matches the particle content used in Ref.~\cite{Grefa:2026meq}, lies below both the lattice-QCD and Bayesian-extracted results, while including $\Lambda$ shifts the proxy result upward, above them. The strangeness-to-electric-charge splitting
ratio $-\Delta\mu_\tS/\Delta\mu_\tQ$ reflects the combined effect of the
electric-charge and strangeness sectors. Its proxy estimates therefore show only moderate sensitivity to the $\Lambda$ contribution, with both proxies lying close to unity and thereby overestimating both the lattice-QCD and Bayesian thermal-analysis results.

\begin{figure}[htbp]
    \centering
    \includegraphics[width=0.9\linewidth]{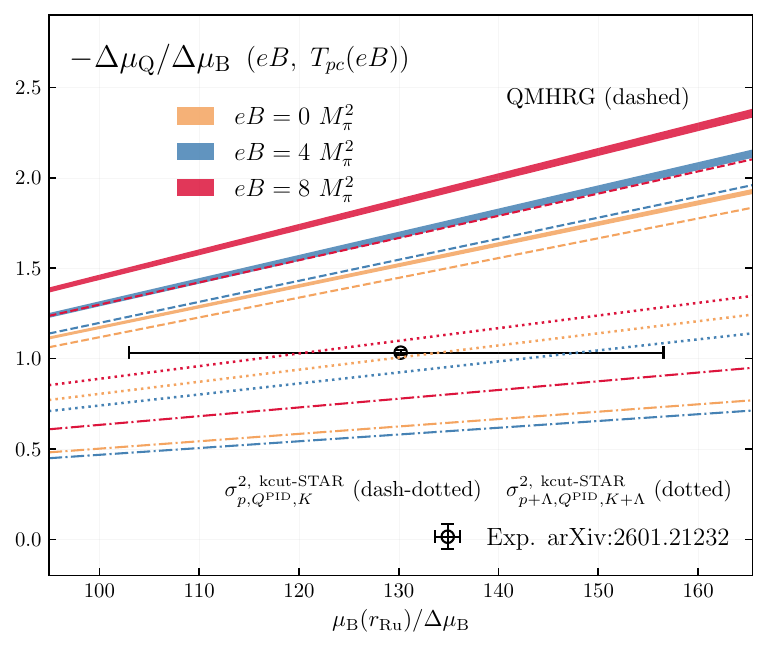}
    
    \includegraphics[width=0.9\linewidth]{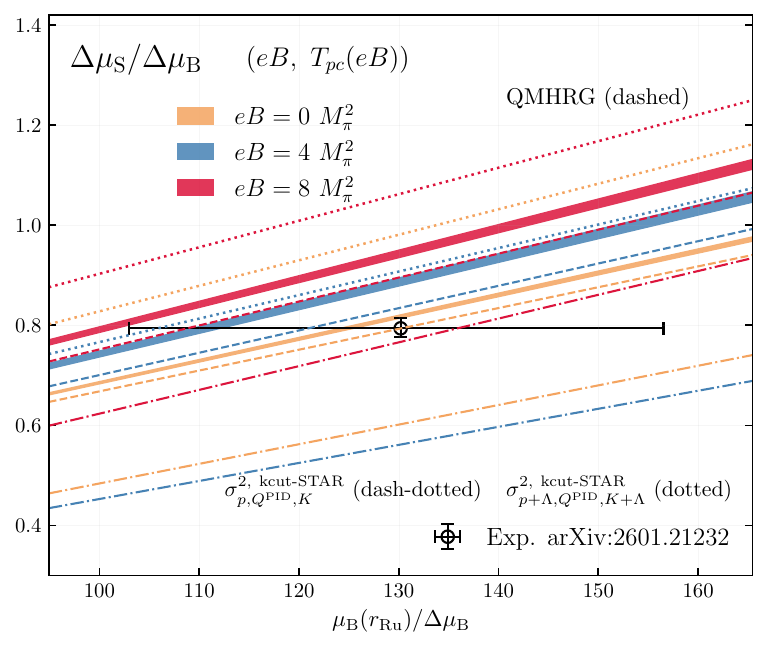}
    \includegraphics[width=0.9\linewidth]{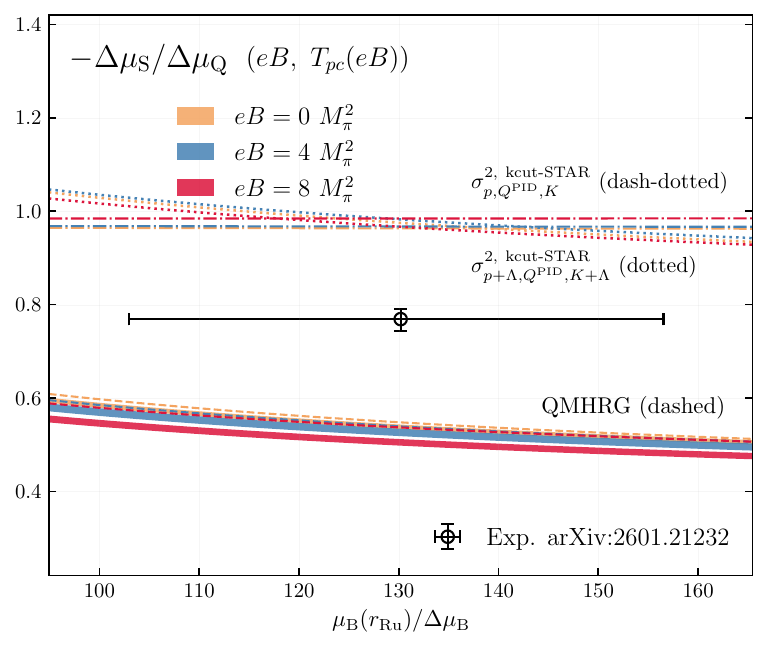}
    
    \caption{Isospin-driven conserved charge splitting ratios, ${-\Delta \mu_\tQ}/{ \Delta \mu_\tB} $, ${\Delta \mu_\tS}/{ \Delta \mu_\tB}$ and ${-\Delta \mu_\tS}/{ \Delta \mu_\tQ} $ (from top to bottom) in presence of magnetic fields along the pseudo-critical line $T_{pc}(eB)$.  Lattice-QCD continuum estimates are represented by colored bands for various magnetic field strengths. The black circles denote Bayesian thermal-analysis extracted ratios and isobar freeze-out chemical potential region. The colored lines denote results within the HRG framework at various magnetic strengths: dashed for full QMHRG results, while the dash-dotted and dotted colored lines correspond to experimental proxies incorporating kinematic cuts relevant to STAR detectors. }
    \label{fig:RdmuBQS_Zr-Ru_Tpc-eB}
\end{figure}

\autoref{fig:RdmuBQS_Zr-Ru_Tpc-eB} extends our lattice-QCD analysis of isospin-driven chemical-potential splittings to nonzero magnetic fields
along the pseudo-critical line $T_{pc}(eB)$. We show the same
sign-adjusted ratios as in \autoref{fig:RdmuBQS_Zr-Ru_T158-eB0},
namely $-\Delta\mu_\tQ/\Delta\mu_\tB$,
$\Delta\mu_\tS/\Delta\mu_\tB$, and
$-\Delta\mu_\tS/\Delta\mu_\tQ$ from top to bottom. The lattice-QCD
continuum estimate bands at $eB=0,\,4,$ and $8~M_\pi^2$ (gold, blue, and red bands, respectively)  show systematic
magnetic-field-induced modifications while preserving the sign structure
discussed above.

In the electric-charge sector, $-\Delta\mu_\tQ/\Delta\mu_\tB$ increases
with $eB$, indicating that $\Delta\mu_\tQ/\Delta\mu_\tB$ becomes more
negative in a stronger magnetic field. The strangeness splitting
$\Delta\mu_\tS/\Delta\mu_\tB$ also increases with $eB$, although the
effect is milder than in the electric-charge sector. These trends are consistent with the magnetic-field response of the leading-order coefficients $q_1(r_{\rm Ru})$ and $s_1(r_{\rm Ru})$ discussed in Sec.~\ref{sec:q1s1_RuZr}. Further details are given in Refs.~\cite{Ding:2025nyh, Ding:2025jfz} in the context of magnetic-field-induced effects. At the same time, the slope with respect to $\mu_\tB(r_{\rm Ru})/\Delta\mu_\tB$ remains largely unaffected by the magnetic field in both sectors, consistent with the weak $eB$ dependence of $q_1(r_{\rm Zr})-q_1(r_{\rm Ru})$ and $s_1(r_{\rm Zr})-s_1(r_{\rm Ru})$.

The bottom panel shows that the strangeness-to-electric-charge splitting
ratio $-\Delta\mu_\tS/\Delta\mu_\tQ$ has only weak magnetic-field
dependence. The bands at $eB=0,\,4,$ and $8~M_\pi^2$ remain close to
each other over the phenomenologically relevant range of
$\mu_\tB(r_{\rm Ru})/\Delta\mu_\tB$, although both
$-\Delta\mu_\tQ/\Delta\mu_\tB$ and
$\Delta\mu_\tS/\Delta\mu_\tB$ increase with $eB$ in the upper two
panels. This indicates that the magnetic-field effects in the
electric-charge and strangeness sectors partly cancel in the ratio. The
magnitude of $\Delta\mu_\tS/\Delta\mu_\tQ$ remains below unity,
confirming that the electric-charge splitting is larger than the
strangeness splitting in the relevant window.

The HRG results at nonzero magnetic fields are also shown as hadronic
reference calculations. In this case, the HRG construction is more
nontrivial than at $eB=0$, because charged hadrons acquire
Landau-quantized spectra in a magnetic field.
The resulting changes in
the density of states and dispersion relations modify the
conserved-charge susceptibilities and thereby the HRG estimates of
$q_1$ and $s_1$ that enter the splitting relations in
Eqs.~(\ref{eq:DmuQ_DmuB})--(\ref{eq:DmuS_DmuQ}). Although the HRG calculations are not expected to be quantitatively reliable at $T_{pc}(eB)$, the ratio observables considered here suppress sensitivity to absolute magnitudes of fluctuations and correlations while retaining the relevant chemical-potential and magnetic-field dependence. The full QMHRG results at nonzero magnetic fields, denoted by dashed colored lines at various magnetic-field strengths $eB$, follow overall trends of the lattice-QCD splitting ratios for all three sectors.

The proxy results shown in \autoref{fig:RdmuBQS_Zr-Ru_Tpc-eB} provide an
experimentally motivated counterpart to the full HRG calculation. Their
construction follows the procedure outlined in Ref.~\cite{Ding:2025jfz},
where proxy observables and kinematic cuts were implemented in the
presence of magnetic fields. 
The STAR-cut proxy, $\sigma^{2,\rm kcut\text{-}STAR}_{p,\,\text{Q}^{\rm PID},\,K}$ constructed from the $\pi,K,p$ proxy set (dash-dotted colored lines), and the corresponding ${\Lambda}$-extended counterparts $\sigma^{2,\rm kcut\text{-}STAR}_{p+ \Lambda,\,\text{Q}^{\rm PID},\,K+\Lambda}$ (dotted colored lines), are shown together with the HRG baselines. At vanishing magnetic field, the inclusion of $\Lambda$ shifts the proxy closer to the Bayesian thermal extraction in the electric-charge sector. In the strangeness sector, the $\Lambda$-extended proxy slightly overshoots the extracted value, while the proxy without $\Lambda$ underestimates it. The mixed ratio $\Delta\mu_\tS/\Delta\mu_\tQ$ shows only moderate sensitivity to $\Lambda$ inclusion. At nonzero magnetic fields, where no corresponding Bayesian thermal-analysis extraction is currently available, both proxies exhibit a mild non-monotonic dependence on $eB$ in all three sectors: the magnitude decreases slightly near $eB\simeq 4\,M_\pi^2$, followed by a shift toward larger values near $eB\simeq 8\,M_\pi^2$.

The moderate magnetic-field dependence of the splitting ratios in
\autoref{fig:RdmuBQS_Zr-Ru_Tpc-eB} can be understood from the structure
of Eqs.~(\ref{eq:DmuQ_DmuB}) and~(\ref{eq:DmuS_DmuB}). As shown in Sec.~\ref{sec:q1s1_RuZr}, the leading-order coefficient $q_1(r)$ has a pronounced magnetic-field dependence, whereas the isospin difference $q_1(r_{\rm Zr})-q_1(r_{\rm Ru})$ changes only mildly with $eB$. 
In the phenomenologically relevant region,
$\mu_\tB(r_{\rm Ru})/\Delta\mu_\tB\simeq 130$,
the isospin-driven difference slope term $\left[q_1(r_{\rm Zr})-q_1(r_{\rm Ru})\right] \mu_\tB(r_{\rm Ru})/\Delta\mu_\tB$ is amplified and  dominates over the intercept term $q_1(r_{\rm Zr})$~\footnote{In Appendix~\ref{app:full_muBoverDmuB}, we provide further details on this matter, exploring the full $\muB(r_{\rm Ru})/\Delta \muB$ window.}
Since this dominant
isospin-difference term has only weak magnetic-field dependence, the
resulting splitting ratios are less sensitive to $eB$ than the
individual coefficient $q_1(r)$ itself. 
This explains why the
magnetic-field enhancement of the isospin-driven splitting ratios is
comparatively modest.

\begin{figure}[!htp]
    \centering

    \includegraphics[width=0.9\linewidth]{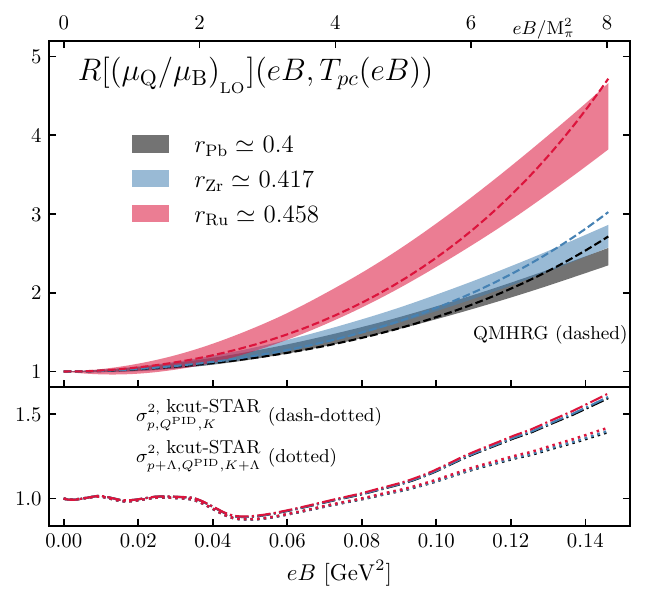}
    
    \includegraphics[width=0.9\linewidth]{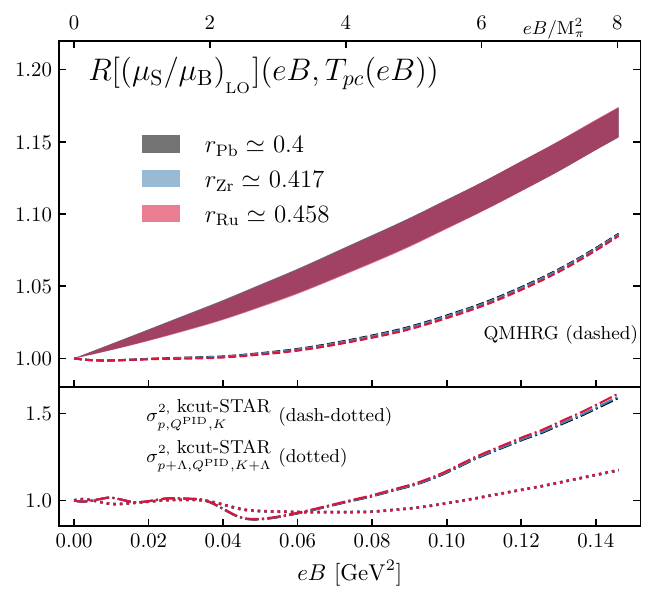}
    
    \includegraphics[width=0.9\linewidth]{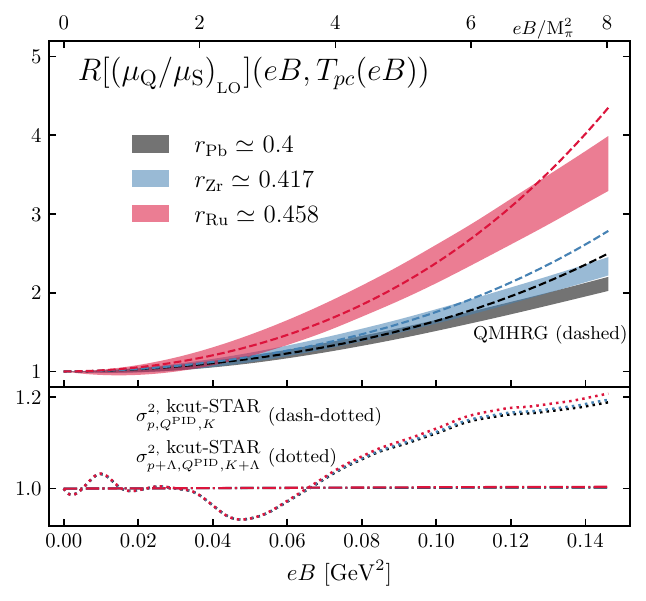}

    \caption{Ratio observables $R[(\mu_\tQ/\mu_\tB)_{\rm LO}]$ (top), $R[(\mu_\tS/\mu_\tB)_{\rm LO}]$ (middle), and $R[(\mu_\tQ/\mu_\tS)_{\rm LO}]$ (bottom) along $T_{pc}(eB)$. Colored bands denote lattice-QCD continuum estimates at fixed isospin parameters, $r_{\rm Pb}\simeq0.4$ (black), $r_{\rm Zr}\simeq0.417$ (blue), and $r_{\rm Ru}\simeq0.458$ (red). Dashed lines denote QMHRG results. Lower sub-panels show proxy ratios with STAR kinematic cuts; dash-dotted for $\sigma^{2,\rm kcut\text{-}STAR}_{p,Q^{\rm PID},K}$, dotted for $\sigma^{2,\rm kcut\text{-}STAR}_{p+\Lambda,Q^{\rm PID},K+\Lambda}$. }
    \label{fig:q1s1_Rcp_PbZrRu_Tpc-eB}
\end{figure}

To construct observables with stronger magnetic-field sensitivity, we
therefore consider normalized response ratios,
$R[\mathcal{O}]=\mathcal{O}\big(eB,T_{pc}(eB)\big)\big/ \mathcal{O}\big(eB=0,T_{pc}(eB=0)\big)$. Such ratios were used in Refs.~\cite{Ding:2023bft,Ding:2025jfz}, where both lattice-QCD and HRG results were presented, to quantify magnetic-field enhancements and to identify $\chi^{\rm BQ}_{11}$ and $q_1 \equiv (\mu_\tQ/\mu_\tB)_{\rm LO}$ as sensitive QCD magnetometers. \autoref{fig:q1s1_Rcp_PbZrRu_Tpc-eB} quantifies these $eB$-induced enhancements of the leading-order coefficients, examining $R[\mathcal{O}]$ for $\mathcal{O} \in \{ (\mu_\tQ/\mu_\tB)_{\rm LO},~(\mu_\tS/\mu_\tB)_{\rm LO},~(\mu_\tQ/\mu_\tS)_{\rm LO} \}$ along $T_{pc}(eB)$, complementary to \autoref{fig:q1s1_PbZrRu_Tpc-eB}. In the top panel, $R[(\mu_\tQ/\mu_\tB)_{\rm LO}]$ exhibits pronounced, scenario-dependent deviations from unity. For Pb$+$Pb, $R[(\mu_\tQ/\mu_\tB)_{\rm LO}] \approx 2.4$ at $eB \simeq 8\,M_\pi^2$. For the isobar systems, $R[(\mu_\tQ/\mu_\tB)_{\rm LO}]$ at $r_{\rm Zr}$ is comparable to the Pb$+$Pb case, while the more isospin-symmetric Ru$+$Ru system responds more strongly, reaching $R[(\mu_\tQ/\mu_\tB)_{\rm LO}] \approx 4$ at $r_{\rm Ru}$ for $eB \simeq 8\,M_\pi^2$. By contrast, the middle panel shows that the lattice-QCD continuum estimates of $R[(\mu_\tS/\mu_\tB)_{\rm LO}]$ have much weaker dependence on $eB$ and negligible dependence on $r$. The continuum
bands for Pb$+$Pb, Zr$+$Zr, and Ru$+$Ru largely overlap, with only $\sim15\%$ enhancement at the largest magnetic fields considered. 
The bottom-panel double ratio,
$R[(\mu_\tQ/\mu_\tS)_{\rm LO}]$, therefore inherits most of its
magnetic-field sensitivity from the electric-charge sector and shows a
qualitatively similar enhancement to
$R[(\mu_\tQ/\mu_\tB)_{\rm LO}]$, although mildly reduced by the small
response of $R[(\mu_\tS/\mu_\tB)_{\rm LO}]$.

The QMHRG results, shown by dashed lines, qualitatively reproduce the
lattice-QCD magnetic-field dependence of
$R[(\mu_\tQ/\mu_\tB)_{\rm LO}]$ in the top panel. In particular, they
capture both the sizable enhancement with increasing $eB$ and the
ordering among the different charge-to-baryon ratios, although
quantitative differences from the lattice-QCD continuum estimates remain
at larger magnetic fields.
For $R[(\mu_\tS/\mu_\tB)_{\rm LO}]$, the QMHRG results are only mildly
different from the lattice-QCD continuum estimates, at the level of a few percent
to about ten percent over the range of $eB$ considered. This moderate
difference is consistent with the bottom panel: since
$R[(\mu_\tQ/\mu_\tS)_{\rm LO}]$ is essentially controlled by the ratio of
the top and middle panels, the QMHRG still gives a reasonable description
of the lattice-QCD trend of the double ratio, although deviations become
more visible at larger $eB$.

The situation changes substantially for the STAR-cut proxy observables
shown in the lower sub-panels. For
$R[(\mu_\tQ/\mu_\tB)_{\rm LO}]$, the proxies still retain a visible
magnetic-field dependence, but they fail to reproduce the pronounced
scenario dependence seen in lattice QCD. In particular, the suppression
relative to the lattice-QCD continuum estimates is most severe for
Ru$+$Ru, where the full lattice-QCD result shows the largest enhancement,
whereas it is less pronounced for Pb$+$Pb and Zr$+$Zr. As a consequence,
the clear separation between the Zr$+$Zr and Ru$+$Ru responses in
lattice QCD is almost completely washed out in the proxy construction.
Including $\Lambda$ further reduces the response and does not restore
the missing isobar sensitivity. For $R[(\mu_\tS/\mu_\tB)_{\rm LO}]$, the $\pi,K,p$ proxy shows an
artificially strong magnetic-field response; with $\chi^{\rm BS}_{11}$ unresolved, it closely resembles the
behavior of $R[(\mu_\tQ/\mu_\tB)_{\rm LO}]$, and therefore overshoots
the much weaker lattice-QCD enhancement. Including $\Lambda$ brings the
proxy result much closer to the lattice-QCD trend, reflecting the
importance of strange baryons for reconstructing the strangeness
chemical-potential response.
 Because the $\pi,K,p$ proxy gives very similar normalized responses in $(\mu_\tQ/\mu_\tB)_{\rm LO}$ and $(\mu_\tS/\mu_\tB)_{\rm LO}$, the
magnetic-field dependence largely cancels in their double ratio  $R[(\mu_\tQ/\mu_\tS)_{\rm LO}]$, leaving it close to unity. Including $\Lambda$ breaks this cancellation and generates a mild enhancement, but the resulting response remains much weaker than the lattice-QCD and full QMHRG results.

Taken together, these proxy results show that the STAR-cut proxies still
retain a visible magnetic-field dependence, with normalized responses
reaching about $1.5$ at the largest magnetic fields considered in some
cases. However, the differential response between the two isobar systems
is largely lost in the proxy construction: the Ru$+$Ru and Zr$+$Zr
proxy results nearly overlap, in contrast to the clear separation
predicted by lattice QCD for
$R[(\mu_\tQ/\mu_\tB)_{\rm LO}]$.
This loss of isobar sensitivity should be attributed mainly to the
restricted particle content of the proxy observables, rather than to the
STAR kinematic cuts themselves. Since the proxies are constructed from
charged, directly detected hadrons, electrically neutral states are
omitted, most importantly the neutron ($B=1$) and the neutral kaon
$K^0$ ($S\neq0$). Their absence affects
$\chi_2^{\rm B}$, $\chi_2^{\rm S}$, $\chi_{11}^{\rm BS}$, and
$\chi_{11}^{\rm QS}$ unequally, while the pion-dominated
$\chi_2^{\rm Q}$ is largely retained. As a result, the leading-order
chemical-potential ratios reconstructed from the proxies respond quite
similarly for $r_{\rm Zr}$ and $r_{\rm Ru}$, causing the Ru--Zr
difference to be largely cancelled even before acceptance cuts are
applied.

\begin{figure}[!htbp]
    \centering
    \includegraphics[width=0.9\linewidth]{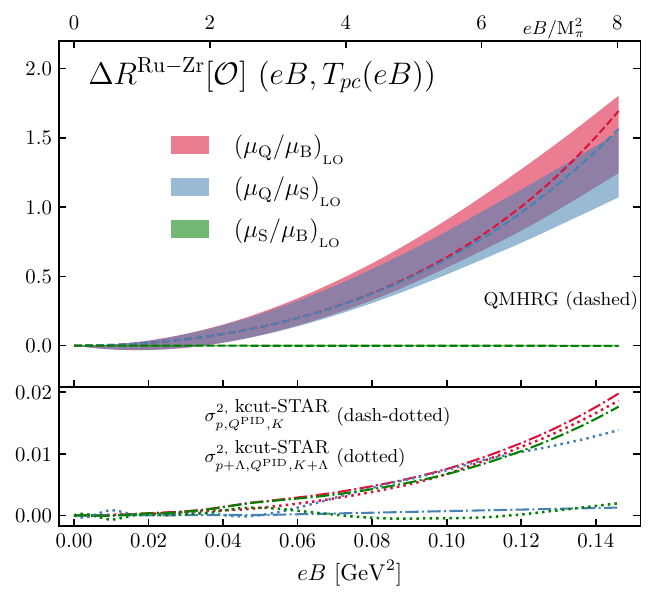}
    \caption{Lattice-QCD continuum estimates for isospin-driven differences of normalized ratios between isobar partners (Zr, Ru), $\Delta R^{\rm Ru-Zr}[\mathcal{O}]$ for $(\mu_\tQ/\mu_\tB)_{\rm LO}$ (red band), $(\mu_\tS/\mu_\tB)_{\rm LO}$ (green band), and $(\mu_\tQ/\mu_\tS)_{\rm LO}$ (blue band) along the pseudo-critical line $T_{pc}(eB)$. Dashed lines denote QMHRG results. The lower sub-panel shows proxy results with STAR kinematic cuts, dash-dotted for $\sigma^{2,\rm kcut\text{-}STAR}_{p,Q^{\rm PID},K}$, dotted for $\Lambda$-extended $\sigma^{2,\rm kcut\text{-}STAR}_{p+\Lambda,Q^{\rm PID},K+\Lambda}$.  }
    \label{fig:DZr-Ru_Rcp_q1s1_Tpc-eB}
\end{figure}

\autoref{fig:DZr-Ru_Rcp_q1s1_Tpc-eB} shows the Ru--Zr differences of
the normalized magnetic-field response ratios along $T_{pc}(eB)$,
providing a more direct view of the isospin dependence already seen in
\autoref{fig:q1s1_Rcp_PbZrRu_Tpc-eB}. 
The largest lattice-QCD signal is found in the electric-charge sector: 
$\Delta R^{\rm Ru-Zr}[(\mu_\tQ/\mu_\tB)_{\rm LO}]$ grows substantially
with $eB$, reflecting the stronger normalized response of Ru$+$Ru than
Zr$+$Zr in $R[(\mu_\tQ/\mu_\tB)_{\rm LO}]$.
The double-ratio difference, 
$\Delta R^{\rm Ru-Zr}[(\mu_\tQ/\mu_\tS)_{\rm LO}]$ shows a
similar but slightly weaker trend, while 
$\Delta R^{\rm Ru-Zr}[(\mu_\tS/\mu_\tB)_{\rm LO}]$ remains close to zero, its band hardly visible owing to the strong Zr–Ru correlation. The full QMHRG results, shown by dashed lines,
follow the same pattern: sizable Ru--Zr differences are obtained in the
electric-charge and double-ratio sectors, whereas the strangeness-to-baryon 
ratio gives an almost vanishing difference.

The lower sub-panel of \autoref{fig:DZr-Ru_Rcp_q1s1_Tpc-eB} shows that
the situation is very different for the STAR-cut proxies. Although the
individual proxy ratios in \autoref{fig:q1s1_Rcp_PbZrRu_Tpc-eB} still
retain a visible overall magnetic-field dependence, their Ru--Zr
differences are tiny: throughout the plotted range they remain at the
percent level, $\Delta R^{\rm Ru-Zr}\lesssim 0.02$. Thus, in the
electric-charge and double-ratio sectors the proxy results are more than
an order of magnitude below the lattice-QCD and full-QMHRG results. This
directly reflects the near overlap of the $r_{\rm Zr}$ and $r_{\rm Ru}$
proxy curves in \autoref{fig:q1s1_Rcp_PbZrRu_Tpc-eB}. As discussed
above, the loss of isospin sensitivity is mainly caused by the restricted
charged-hadron content of the proxy construction, in particular the
omission of neutral states such as the neutron and $K^0$. 
Therefore, while lattice QCD predicts
a sizable differential magnetic-field response between the two isobar
systems, this Ru--Zr sensitivity is largely washed out in the
charged-hadron proxy observables.

\section{Conclusions}
\label{sec:summary}

In this work, we have presented first-principles $(2+1)$-flavor lattice-QCD results for isospin-driven splittings of conserved-charge chemical potentials in the QCD crossover region, both at vanishing and strong magnetic fields along the pseudo-critical line $T_{pc}(eB)$. We have outlined a framework incorporating strangeness neutrality and    
charge-to-baryon ratio $r$ that maps the isospin asymmetry encoded in $r_{\rm Zr}$ and $r_{\rm Ru}$ onto the experimentally accessible ratios of chemical-potential
splittings, $\Delta\mu_\tQ/\Delta\mu_\tB$, $\Delta\mu_\tS/\Delta\mu_\tB$, and $\Delta\mu_\tS/\Delta\mu_\tQ$, expressed as functions of $\mu_\tB(r_{\rm
Ru})/\Delta\mu_\tB$.

The splitting ratios are controlled by the leading-order coefficients
$q_1\equiv(\mu_\tQ/\mu_\tB)_{\rm LO}$ and
$s_1\equiv(\mu_\tS/\mu_\tB)_{\rm LO}$, together with their
isospin-driven differences between the Zr+Zr and Ru+Ru systems. Our
lattice-QCD continuum estimates show a clear hierarchy in the
electric-charge sector: for $r<0.5$, $q_1<0$ and
$|q_1(r_{\rm Zr})|>|q_1(r_{\rm Ru})|$, implying
$q_1(r_{\rm Zr})-q_1(r_{\rm Ru})<0$. Magnetic fields enhance the
magnitude of $q_1$, whereas the strangeness coefficient $s_1$ remains
positive and exhibits much milder isospin and magnetic-field dependence.

Near the isobar freeze-out regime at $T=158^{+4}_{-4}~{\rm MeV}$ and $eB=0$, we find $\Delta\mu_\tQ/\Delta\mu_\tB<0$, $\Delta\mu_\tS/\Delta\mu_\tB>0$ and $\Delta\mu_\tS/\Delta\mu_\tQ<0$. The magnitude of the electric-charge
splitting grows faster with $\mu_\tB(r_{\rm Ru})/\Delta\mu_\tB$ than the
strangeness splitting, so the electric-charge sector dominates the
strangeness-to-electric-charge splitting ratio. Comparison with the recent Bayesian thermal-analysis extraction from isobar-collision hadron yields shows the same sign structure and comparable magnitudes, with $\Delta\mu_\tS/\Delta\mu_\tB$ showing the closest agreement with the
lattice-QCD continuum estimate. 

Extending to strong magnetic fields along the pseudo-critical line $T_{pc}(eB)$, the magnitudes of $-\Delta\mu_\tQ/\Delta\mu_\tB$ and
$\Delta\mu_\tS/\Delta\mu_\tB$ increase with $eB$, while their slopes with
respect to $\mu_\tB(r_{\rm Ru})/\Delta\mu_\tB$ change only mildly. This reflects the fact that, in the phenomenologically relevant region, the amplified isospin-difference term  $\left[q_1(r_{\rm Zr})-q_1(r_{\rm Ru})\right]\mu_\tB(r_{\rm Ru})/\Delta\mu_\tB$ dominates the splitting, but has only weak $eB$-dependence. Consequently, the splitting ratios themselves show only moderate magnetic-field sensitivity. These results are supplemented by HRG calculations and experimentally motivated proxy observables with STAR kinematic cuts. These hadronic reference results show that particle proxies and
acceptance cuts can visibly affect the inferred splitting ratios,
especially in the electric-charge sector and in observables involving
baryon--strangeness correlations.

To further diagnose magnetic-field effects in the conserved-charge
sector, we examined Ru--Zr differences of normalized magnetic-field response ratios,
$\Delta R^{\rm Ru-Zr}[\mathcal{O}]$, for
$\mathcal{O}\in\{(\mu_\tQ/\mu_\tB)_{\rm LO},
(\mu_\tS/\mu_\tB)_{\rm LO},(\mu_\tQ/\mu_\tS)_{\rm LO}\}$. 
In the lattice-QCD continuum estimates, the strongest
response is found for
$\Delta R^{\rm Ru-Zr}[(\mu_\tQ/\mu_\tB)_{\rm LO}]$, indicating that the
relative magnetic-field enhancement of the electric-charge chemical
potential is stronger in Ru+Ru than in Zr+Zr. In contrast,
$\Delta R^{\rm Ru-Zr}[(\mu_\tS/\mu_\tB)_{\rm LO}]$ remains close to zero
within uncertainties.
Thus, $\mu_\tQ$-related normalized ratios provide
the clearest first-principles diagnostic of the isospin-dependent
magnetic-field response in the full conserved-charge sector. In the
STAR-cut proxies, however, this differential response is largely washed
out: the proxy results for $r_{\rm Zr}$ and $r_{\rm Ru}$ nearly overlap,
leaving percent-level Ru--Zr differences that are more than an order of
magnitude below the lattice-QCD and full-QMHRG results in the
electric-charge and double-ratio sectors. This mainly reflects the
restricted charged-hadron content of the proxy construction, especially
the omission of neutral states such as the neutron and $K^0$.

Overall, the consistency in sign and magnitude between the lattice-QCD
results at vanishing magnetic fields and the Bayesian thermal-analysis extraction supports the
isospin-driven splitting framework and highlights isospin as a useful
thermodynamic handle on conserved-charge chemical potentials. At nonzero magnetic fields, the splitting ratios themselves show only
moderate sensitivity, while the normalized Ru--Zr response differences
reveal a stronger differential response in the full lattice-QCD
conserved-charge observables. Our results therefore provide first-principles benchmarks for the
underlying conserved-charge response to isospin and magnetic fields in
isobar collisions, while clarifying the limitations of present proxy
constructions. Further progress may benefit from proxy observables or experimental
combinations that retain more of the full conserved-charge content,
together with lattice-QCD studies of their temperature and
magnetic-field dependence.

\section*{Acknowledgments}
This work is supported partly by the National Natural Science Foundation of China under Grants No. 12293064, No. 12293060, and No. 12325508, as well as the National Key Research and Development Program of China under Contract No. 2022YFA1604900 and the Fundamental Research Funds for the Central Universities, Central China Normal University under Grants No. 30101250314 and No.30106250152. The numerical simulations have been performed on the GPU cluster in the Nuclear Science Computing Center at Central China Normal University ($\mathrm{NSC}^{3}$) and Wuhan Supercomputing Center.

\bibliographystyle{JHEP.bst}
\bibliography{refs.bib}

\appendix

\begin{figure*}[!thp]
    \centering
    \includegraphics[width=0.31\linewidth]{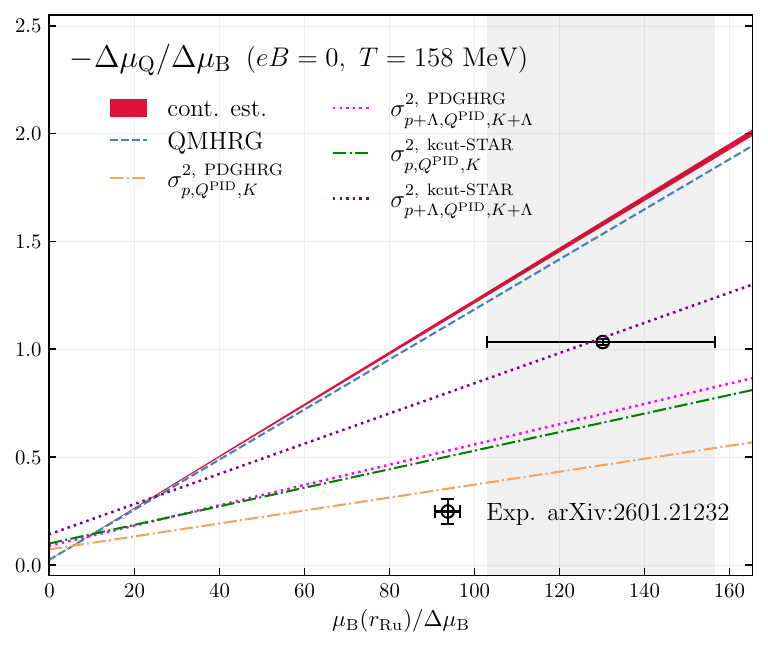}
    \includegraphics[width=0.31\linewidth]{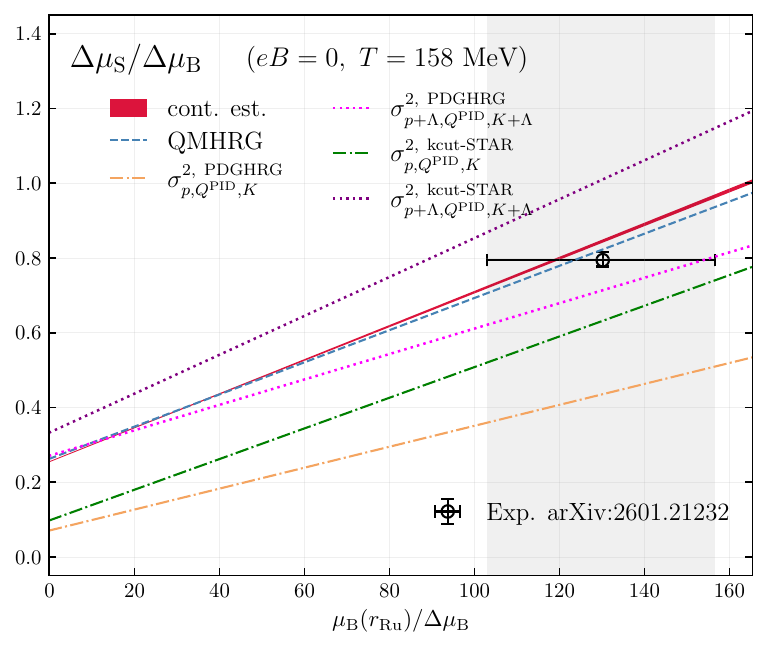}
    \includegraphics[width=0.31\linewidth]{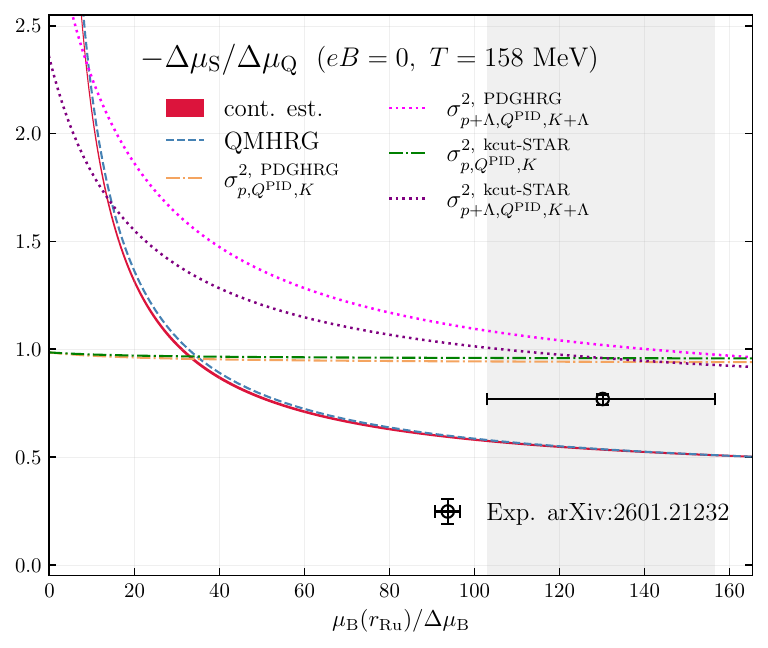}
    
    \caption{Isospin-driven conserved charge splitting ratios, ${-\Delta \mu_\tQ}/{ \Delta \mu_\tB} $, ${\Delta \mu_\tS}/{ \Delta \mu_\tB}$ and ${-\Delta \mu_\tS}/{ \Delta \mu_\tQ} $ (from left to right) near the isobar freeze-out $T=158~{\rm MeV}$. Same as \autoref{fig:RdmuBQS_Zr-Ru_T158-eB0}, but in an extended $\muB(r_{\rm Ru})/\Delta \muB$ window and incorporating experimental proxies constructed using $\pi,~K,~p$ and extended-$\Lambda$ set denoted by dash-dotted and dotted colored lines, respectively.}
    \label{fig:RdmuBQS_Zr-Ru_T158-eB0_full}
\end{figure*}

\section{Next-to-leading-order correction to chemical potential splitting ratios}
\label{app:NLO}

The isospin-driven chemical-potential splitting framework of Sec.~\ref{sec:isospin-driven-framework} is built with the leading-order coefficients $q_1$ and $s_1$. Next-to-leading-order (NLO) corrections follow from incorporating $q_3$ and $s_3$ in Eqs.~\ref{eq:muQ} and~\ref{eq:muS};  throughout, the subscript "NLO" denotes the cumulative result including the leading order. Explicit expressions for $q_3$ and $s_3$, given in Ref.~\cite{Bazavov:2017dus}, involve 4th-order fluctuations and correlations of conserved charges. The chemical-potential splittings between Zr and Ru systems in Eq.~\ref{eq:DmuX} then receive higher-order contributions: 
\bea
(\Delta\hmu_\tQ)_{\rm NLO} \equiv& q_{1}(r_{\rm Zr}) \hmu_\tB(r_{\rm Zr})  -q_{1}(r_{\rm Ru}) \hmu_\tB(r_{\rm Ru}) \nn \\ 
&+~q_{3}(r_{\rm Zr}) \hmu_\tB^3(r_{\rm Zr})- q_{3}(r_{\rm Ru}) \hmu_\tB^3(r_{\rm Ru})
\eea
and similarly for the strangeness sector $(\Delta\hmu_\tS)_{\rm NLO}$ involving $s_1,~s_3$. In Ref.~\cite{Ding:2023bft}, some of the current authors had computed $q_3/q_1$ corresponding to Zr and Ru systems and found that the resulting ratio remains within $~2\%$ and even decreases with magnetic field strength. \footnote{These computations were performed on $N_\tau=8$ lattices; since $q_3/q_1$ shows no significant discretization effects at $eB=0$~\cite{Bazavov:2012vg,Bazavov:2017dus}, NLO corrections are expected to remain mild in continuum.}

Substituting $\hmu_\tB(r_{\rm Zr})$ as in Sec.~\ref{sec:isospin-driven-framework}, the NLO splitting ratio reads 
\bea
\label{eq:DmuQ_DmuB_NLO}
\left(\frac{\Delta\mu_\tQ}{\Delta\mu_\tB}\right)_{\rm NLO} \equiv& q_{1}(r_{\rm Zr}) +  \left( q_{1}(r_{\rm Zr}) -q_{1}(r_{\rm Ru})\right)\frac{\mu_\tB(r_{\rm Ru})}{\Delta\mu_\tB}\nn \\
&+~\left(q_{3}(r_{\rm Zr}) - q_{3}(r_{\rm Ru})\right) \hmu_\tB^2(r_{\rm Ru})  (\frac{\mu_\tB(r_{\rm Ru})}{\Delta\mu_\tB}) \nn \\
&+~3q_{3}(r_{\rm Zr})\hmu_\tB^2(r_{\rm Ru}) , 
\eea
where terms of $\mathcal{O}\big((\Delta\hmu_\tB)^2\big)$ and higher have been dropped, as they are negligible compared to the NLO corrections kept above. Both NLO correction terms in Eq.~(\ref{eq:DmuQ_DmuB_NLO}) carry an overall factor $\hmu_\tB^2(r_{\rm Ru})\equiv\big(\mu_\tB(r_{\rm Ru})/T\big)^2$ from the Taylor expansion in $\hmu_\tB$. Near the isobar freeze-out point, $\mu_\tB(r_{\rm Ru})\simeq 20.31~{\rm MeV}$ and $T\simeq158~{\rm MeV}$ give $\hmu_\tB^2(r_{\rm Ru})\approx 2\% \sim\mathcal{O}(10^{-2})$. The lattice estimate $q_3/q_1\lesssim 2\%\sim\mathcal{O}(10^{-2})$~\cite{Ding:2023bft} then suppresses the NLO corrections to $\Delta\mu_\tQ/\Delta\mu_\tB$ to $\mathcal{O}(10^{-3})$ relative to the leading order, well below current lattice uncertainties. Similar arguments extend to the strangeness sector $\Delta\mu_\tS/\Delta\mu_\tB$ with coefficients $s_1,~s_3$, and to the mixed ratio $\Delta\mu_\tS/\Delta\mu_\tQ$. The leading-order splitting-ratio framework of Sec.~\ref{sec:isospin-driven-framework} therefore captures the chemical-potential splittings to sub-percent accuracy near the isobar freeze-out point.


\section{Results in an extended $\muB(r_{\rm Ru})/\Delta \muB$ window}
\label{app:full_muBoverDmuB}
\autoref{fig:RdmuBQS_Zr-Ru_T158-eB0_full} presents the splitting ratios near isobar freeze-out $T=158~{\rm MeV}$ and $eB=0$ over an extended $\mu_\tB(r_{\rm Ru})/\Delta\mu_\tB$ window that spans from vanishing values through the experimental regime (shaded region), complementing \autoref{fig:RdmuBQS_Zr-Ru_T158-eB0}. For the electric-charge and strangeness sectors, the splitting ratios $-\Delta\mu_\tQ/\Delta\mu_\tB$ and $\Delta\mu_\tS/\Delta\mu_\tB$ (left and middle panels) keep increasing linearly with $\mu_\tB(r_{\rm Ru})/\Delta\mu_\tB$, and lattice uncertainties, although relatively small, broaden as isospin-driven differences dominate, especially near the experimental window. By contrast, the mixed ratio $-\Delta\mu_\tS/\Delta\mu_\tQ$ (right panel) in the experimental window begins to saturate to $\sim 1/2$, as the $\mu_\tB(r_{\rm Ru})/\Delta\mu_\tB$ dependence eventually cancels. This saturation value follows from the bottom panel of \autoref{fig:q1s1_PbZrRu_Tpc-eB}, where the isospin-driven difference in the strangeness sector is suppressed by a factor of $\sim 2$ relative to the electric-charge sector.

\autoref{fig:RdmuBQS_Zr-Ru_T158-eB0_full} also shows the corresponding PDGHRG proxy results as colored broken lines: dash-dotted for $\sigma^{2,\rm PDGHRG}_{p,\tQ^{\rm PID},K}$ and dotted for the $\Lambda$-extended set $\sigma^{2,\rm PDGHRG}_{p+\Lambda,\tQ^{\rm PID},K+\Lambda}$. These results lay the foundation for the kinematic-cut proxies in \autoref{fig:RdmuBQS_Zr-Ru_T158-eB0}. The $\Lambda$-extended set substantially improves the description of the $\chi^{\rm BS}_{11}$ sector. This is clearly visible in the small-$\mu_\tB(r_{\rm Ru})/\Delta\mu_\tB$ window of the strangeness-sector splittings, where $\sigma^{2,\rm PDGHRG}_{ p+\Lambda,\tQ^{\rm PID},K+\Lambda}$ shifts closer to the lattice and HRG results. In the mixed ratio $-\Delta\mu_\tS/\Delta\mu_\tQ$, the standard $\pi,~K,~p$ set, both with and without kinematic cuts, stays near unity across the full $\mu_\tB(r_{\rm Ru})/\Delta\mu_\tB$ window, while the extended proxy gradually decreases initially, beginning to saturate near unity in the experimental window.

\end{document}